\pdfoutput=1
\documentclass{article}

\usepackage{arxiv}

\usepackage[utf8]{inputenc} 
\usepackage[T1]{fontenc}    
\usepackage{lmodern}        
\usepackage{hyperref}       
\usepackage{url}            
\usepackage{booktabs}       
\usepackage{amsfonts}       
\usepackage{nicefrac}       
\usepackage{microtype}      
\usepackage{graphicx}

\title{\pkg{ergm} 4: Computational Improvements}

\author{
    Pavel N. Krivitsky
   \\
    School of Mathematics and Statistics \\
    University of New South Wales \\
  Sydney, NSW, Australia \\
  \texttt{\href{mailto:p.krivitsky@unsw.edu.au}{\nolinkurl{p.krivitsky@unsw.edu.au}}} \\
   \And
    David R. Hunter
   \\
    Department of Statistics \\
    Penn State University \\
  State College, PA, USA \\
  \texttt{\href{mailto:dhunter@stat.psu.edu}{\nolinkurl{dhunter@stat.psu.edu}}} \\
   \And
    Martina Morris
   \\
    Departments of Sociology and Statistics \\
    University of Washington \\
  Seattle, WA, USA \\
  \texttt{\href{mailto:morrism@uw.edu}{\nolinkurl{morrism@uw.edu}}} \\
   \And
    Chad Klumb
   \\
    Center for Studies in Demography and Ecology \\
    University of Washington \\
  Seattle, WA, USA \\
  \texttt{\href{mailto:cklumb@uw.edu}{\nolinkurl{cklumb@uw.edu}}} \\
  }

\usepackage{color}
\usepackage{fancyvrb}

\DefineVerbatimEnvironment{Highlighting}{Verbatim}{commandchars=\\\{\}}
\usepackage{framed}
\definecolor{shadecolor}{RGB}{248,248,248}
\newenvironment{Shaded}{\begin{snugshade}}{\end{snugshade}}

\newcommand{\CharTok}[1]{\textcolor[rgb]{0.31,0.60,0.02}{#1}}
\newcommand{\CommentTok}[1]{\textcolor[rgb]{0.56,0.35,0.01}{\textit{#1}}}

\newcommand{\ControlFlowTok}[1]{\textcolor[rgb]{0.13,0.29,0.53}{\textbf{#1}}}
\newcommand{\DataTypeTok}[1]{\textcolor[rgb]{0.13,0.29,0.53}{#1}}
\newcommand{\DecValTok}[1]{\textcolor[rgb]{0.00,0.00,0.81}{#1}}

\newcommand{\ErrorTok}[1]{\textcolor[rgb]{0.64,0.00,0.00}{\textbf{#1}}}

\newcommand{\FloatTok}[1]{\textcolor[rgb]{0.00,0.00,0.81}{#1}}

\newcommand{\KeywordTok}[1]{\textcolor[rgb]{0.13,0.29,0.53}{\textbf{#1}}}
\newcommand{\NormalTok}[1]{#1}
\newcommand{\OperatorTok}[1]{\textcolor[rgb]{0.81,0.36,0.00}{\textbf{#1}}}
\newcommand{\OtherTok}[1]{\textcolor[rgb]{0.56,0.35,0.01}{#1}}

\newcommand{\StringTok}[1]{\textcolor[rgb]{0.31,0.60,0.02}{#1}}

\providecommand{\tightlist}{%
  \setlength{\itemsep}{0pt}\setlength{\parskip}{0pt}}

\newlength{\cslhangindent}
\newlength{\csllabelwidth}
\newlength{\cslentryspacingunit} 
\newenvironment{cslreferences}%
  {\setlength{\parindent}{0pt}%
  \everypar{\setlength{\hangindent}{\cslhangindent}}\ignorespaces}%
  {\par}
 {
  \setlength{\parindent}{0pt}
  \ifodd #1
  \let\oldpar\par
  \def\par{\hangindent=\cslhangindent\oldpar}
  \fi
  \setlength{\parskip}{#2\cslentryspacingunit}
 }%
 {}
\usepackage{calc}

\usepackage{ifthen}
\usepackage{amsmath,amssymb,tikz,enumitem,bm,booktabs,etoolbox,array}
\setlist{noitemsep}
\usepackage[font=normalsize]{subfig}

\AtBeginEnvironment{Highlighting}{\footnotesize}
\makeatletter
\renewcommand{\verbatim@font}{\ttfamily\footnotesize}
\makeatother
\def\y{\mathbf{y}}
\def\Y{\mathbf{Y}}

\def\cnmapel{\eta}
\def\cnmap{\bm{\cnmapel}}
\def\dcnmap{\cnmap'}
\def\nnatpar{ p }
\def\curvparel{\theta}
\def\curvpar{{\bm{\curvparel}}}

\def\natcurvpars{\bm{\Theta}_{\text{N}}}
\def\mle{\hat{\curvpar}}
\def\mple{\tilde{\curvpar}}

\def\ncurvpar{ q }
\def\genstatel{g}
\def\genstats{\mathbf{\genstatel}}

\def\diffstatel{z}
\def\diffstats{\bm{\diffstatel}}

\def\proposed{^{\text{proposed}}}
\def\current{^{\text{current}}}

\def\sampind{ s }
\def\guessind{ t }
\def\sampsize{ S }

\newcommand\sumsamp[1][]{ \sum_{\sampind#1=1}^{\sampsize#1} }
\def\meansamp{\frac{1}{\sampsize}\sumsamp}

\def\changeij{\bm{\Delta}\sij}

\def\grad{\bm{\nabla}}

\DeclareMathOperator{\E}{E}
\DeclareMathOperator{\Var}{Var}
\DeclareMathOperator{\logit}{logit}

\DeclareMathOperator{\ERGM}{ERGM}

\DeclareMathOperator{\0}{\mathbf{0}}

\def\netsY{\mathcal{Y}}

\DeclareMathOperator{\Uniform}{Unif}

\DeclareMathOperator{\tr}{tr}
\DeclareMathOperator{\Prob}{Pr}

\DeclareMathOperator{\modop}{mod}
\def\llik{\ell}

\def\h{h}

\def\normc{\kappa}

\def\energy{E}

\def\cheg{\normc_{\h,\cnmap,\genstats}}

\def\Ptyheg{\Prob_{\curvpar,\netsY,\h,\cnmap,\genstats}}

\def\Dyheg{\ERGM_{\netsY,\h,\cnmap,\genstats}}

\newcommand{\DY}[1]{\ERGM_{#1}}

\newcommand{\PTyheg}[1]{\Prob_{#1;\netsY,\h,\cnmap,\genstats}}
\newcommand{\ETyheg}[1]{\E_{#1;\netsY,\h,\cnmap,\genstats}}
\newcommand{\VTyheg}[1]{\Var_{#1;\netsY,\h,\cnmap,\genstats}}

\def\argmax{\arg\max}

\def\reals{\mathbb{R}}

\def\pij{{(i,j)}}

\def\ynetsY{{\y\in\netsY}}
\def\ypnetsY{{\y'\in\netsY}}
\def\sij{_{i,j}}

\def\Yij{Y\!\sij}
\def\yij{y\sij}

\def\Yy{\Y=\y}
\def\sobs{^{\text{obs}}}

\def\yobs{\y\sobs}

\def\guess{\curvpar^\guessind}

\def\nextguess{\curvpar^{\guessind+1}}

\newcommand{\natpar}[1][]{\cnmap#1(\curvpar)}
\newcommand{\natparT}[2][]{\cnmap#1(\curvpar^{#2})}
\newcommand{\natparp}[1][]{\cnmap#1(\curvpar')}

\newcommand{\dnatpar}[1][]{\dcnmap#1(\curvpar)}

\providecommand{\abs}[1]{\lvert#1\rvert}

\def\t{^\top}

\def\defeq{\stackrel{\text{def}}{=}}

\newcommand{\innerprod}[2]{#1\t#2}

\def\beq{\begin{equation}}
\def\eeq{\end{equation}}

\newcommand{\ENBC}[1]{\left\{ #1 \right\}}
\newcommand{\en}[3]{#1 #3 #2}
\newcommand{\enbc}[1]{\{ #1 \}}

\providecommand{\E}{\operatorname{E}}

\providecommand{\VAR}{\operatorname{var}}
\def\Var{\VAR}

\newcommand{\figref}[1]{Figure~\ref{#1}}
\newcommand{\secref}[1]{Section~\ref{#1}}
\newcommand{\tabref}[1]{Table~\ref{#1}}

\def\given{\mathop{\,|\,}\nolimits}

\providecommand\pkg[1]{\textbf{#1}}
\providecommand\proglang[1]{\textsf{#1}}
\providecommand\code[1]{\texttt{#1}}

\def\statnet{\pkg{statnet}}

\begin{document}
\maketitle

\begin{abstract}
The \pkg{ergm} package supports the statistical analysis and simulation
of network data. It anchors the \statnet{} suite of packages for network
analysis in \proglang{R} introduced in a special issue in \emph{Journal
of Statistical Software} in 2008. This article provides an overview of
the performance improvements in the 2021 release of \pkg{ergm} version
4. These include performance enhancements to the Markov chain Monte
Carlo and maximum likelihood estimation algorithms as well as broader
and faster searching for networks with certain target statistics using
simulated annealing.
\end{abstract}


\hypertarget{introduction}{%
\section{Introduction}\label{introduction}}

\label{sec:Introduction}

The \statnet{} suite of packages for \proglang{R} (\proglang{R} Core
Team, 2021) was first introduced in 2008, in volume 24 of \emph{Journal
of Statistical Software}, a special issue devoted to \statnet{}.
Together, these packages, which had already gone through the maturing
process of multiple releases, provided an integrated framework for the
statistical analysis of network data: from data storage and
manipulation, to visualization, estimation and simulation. Since that
time the existing packages have undergone continual updates to improve
and add capabilities, and many new packages have been added to extend
the range of network data that can be modeled (e.g., dynamic, valued,
sampled, multilevel). It is the \pkg{ergm} package, however, that
provides the statistical foundation for all of the other modeling
packages in the \statnet{} suite. Version 4 of \pkg{ergm}, released in
2021, is a major upgrade, representing more than a decade of changes and
improvements since Hunter et al. (2008). This paper describes updates to
the central MCMC and SAN algorithms in the package, demonstrating that
these changes have produced substantial improvement in computational
speed and efficiency. It is a companion to Krivitsky, Hunter, et al.
(2022), which discusses improvements to the user interface and modelling
capabilities of \pkg{ergm}.

We repeat the brief presentation from Krivitsky, Hunter, et al. (2022)
of the fully general ERGM framework, referring interested readers to
Schweinberger et al. (2020) for additional technical details. A random
network \(\Y\) is distributed according to an ERGM, written
\(\Y\sim\Dyheg(\curvpar)\), if \begin{equation}\label{ergm}
\Ptyheg(\Yy) = \frac{\h(\y)\exp\en\{\}{\natpar\t\genstats(\y)}} {\cheg(\curvpar,\netsY)},\ \ynetsY.
\end{equation} In Equation \eqref{ergm}, \(\netsY\) is the sample space
of networks; \(\curvpar\) is a \(\ncurvpar\)-dimensional parameter
vector; \(\h(\y)\) is a reference measure, typically a constant in the
case of binary ERGMs; \(\cnmap\) is a mapping from \(\curvpar\) to the
\(\nnatpar\)-vector of canonical parameters, given by the identity
mapping in non-curved ERGMs; \(\genstats\) is a \(\nnatpar\)-vector of
sufficient statistics; and \(\cheg(\curvpar,\netsY)\) is the normalizer
given by
\(\sum_{\ypnetsY} \h(\y') \exp\en\{\} {\natpar\t\genstats(\y')}\), which
is often intractable for models that seek to reproduce the dependence
across ties induced by social effects such as triadic closure. The
\emph{natural parameter space} of the model is
\(\natcurvpars \defeq \{\curvpar: \cheg(\curvpar,\netsY) < \infty\}\).

In the examples throughout the paper we assume that the reader is
familiar with the basic syntax and features of \pkg{ergm} included in
the 2008 \emph{JoSS} volume. Where possible we demonstrate new, more
general, functionality by comparison, using the old syntax and the new
to produce the same result, then moving on with the new syntax to
demonstrate the additional utilities.

The source code for the latest development version of the \pkg{ergm}
package, along with the \texttt{LICENSE} information under GPL-3, is
available at \url{https://github.com/statnet/ergm}.

\hypertarget{markov-chain-monte-carlo-enhancements}{%
\section{Markov chain Monte Carlo
enhancements}\label{markov-chain-monte-carlo-enhancements}}

\label{sec:MCMCEnhancements}

The simulation of random networks distributed according to a particular
ERGM with a known value of the parameter \(\cnmap\) is central to nearly
all functionality of the \pkg{ergm} package. Clearly, simulation is
useful to examine population characteristics of an ERGM using Monte
Carlo methods; potentially less obvious is the role that simulation
plays in the process of maximum likelihood estimation itself. Markov
chain Monte Carlo (MCMC) methods are the means by which \pkg{ergm}
implements simulation of networks, and these methods are therefore the
workhorses of the package.

\hypertarget{summary-of-metropolishastings-algorithms}{%
\subsection{Summary of Metropolis--Hastings
algorithms}\label{summary-of-metropolishastings-algorithms}}

As explained by Hunter et al. (2008), the goal of MCMC is to create a
Markov chain whose stationary distribution is exactly equal to the ERGM
with a given value of \(\cnmap\). After letting the chain run for a long
time, its state may be taken to be an approximate draw from the ERGM in
question. The \pkg{ergm} package does this via a Metropolis--Hastings
algorithm, a special case of MCMC in which at each iteration, a move
from the current network to a new network is proposed according to some
probability distribution. The M--H algorithm operates by allowing only
two possibilities following this proposal: Either the chain remains at
the current network for the next iteration, or the proposed network
becomes the current network for the next iteration. The latter
possibility occurs with probability
\begin{equation}\label{metropolishastings}
\min \left\{ 1 ,
\frac { \Ptyheg(\Yy\proposed) }
{ \Ptyheg(\Yy\current) }
\times \frac { q(\y\current \given \y\proposed) }
{ q(\y\proposed\given \y\current) }
\right\},
\end{equation} where \(q\) denotes the proposal distribution; more
specifically, \(q({\bf a} \given {\bf b})\) is the probability of
proposing \({\bf a}\) if the current state is \({\bf b}\).

It is useful to introduce a \emph{change statistic} or \emph{change
score} \begin{equation}
\changeij\genstats(\y)\defeq\genstats\en[]{\y\cup\enbc{\pij}}-\genstats\en[]{\y\setminus\enbc{\pij}},
\label{eq:changestat}
\end{equation} the effect on the vector of statistics if one were to
change the state of the \((i,j)\) relationship from 0 to 1 while holding
the rest of the network \(\y\) fixed. Let us assume for now that \(q\)
only allows for changing, or toggling, at most one dyad, which is to say
that \(q({\bf a} \given {\bf b})\) must be zero whenever \({\bf a}\) and
\({\bf b}\) differ by more than a single edge. If we call the ratio in
Expression \eqref{metropolishastings} the ``acceptance ratio,'' or AR,
then for the proposed toggle of dyad \(\yij\),
\begin{equation}\label{singletoggle}
\log \text{AR} =
\pm\cnmap\t
\changeij\genstats(\y) + \log
\frac { q(\y\current \given \y\proposed) }
{ q(\y\proposed\given \y\current) },
\end{equation} where the sign in front of
\(\cnmap\t\changeij\genstats(\y)\) is positive when \(\yij\proposed=1\)
and negative when \(\yij\proposed=0\).

It is useful to consider a couple of special cases of the
Metropolis--Hastings algorithm \eqref{metropolishastings}. When we
define \(q(\y\proposed\given \y\current)\) to be proportional to
\(\Ptyheg(\Yy\proposed)\), the value of AR is always 1, which implies
that the proposal is always accepted and the resulting algorithm is
called full-conditional Gibbs sampling. Another special case is the
symmetric proposal, in which
\(q({\bf a} \given {\bf b})=q({\bf b} \given {\bf a})\) for all
\({\bf a}\) and \({\bf b}\), in which case \(\log \mbox{AR}\) is simply
\(\pm\cnmap\changeij\genstats(\y)\). In particular, perhaps the most
basic network-based Metropolis--Hastings algorithm for binary networks
with \(N\) possible edges operates by selecting \(\y\proposed\)
uniformly from among all \(N\) networks that differ from \(\y\current\)
by exactly one edge toggle; thus, \(q(\y\proposed\given \y\current)\)
equals \(N^{-1}\) or 0, depending on whether or not \(\y\proposed\) and
\(\y\current\) differ by exactly one toggle.

Simulation of networks from an ERGM using MCMC can be done using
\texttt{simulate}, documented at \texttt{?simulate.ergm}. In place of
its original \texttt{statsonly=} argument, \texttt{simulate} methods now
take a more versatile \texttt{output} argument, which defaults to
\texttt{"network"} for returning a list of generated \texttt{network}
objects, \texttt{"stats"} for network statistics, \texttt{"edgelist"}
for a more compact representation of the network, or a user-defined
function to be evaluated on the sampler state and returned. For example,
the following code produces triangle counts for fifty random undirected
10-node networks where each edge occurs independently with probability
2/3. Also, the model statistics---in this case, edge counts---are
attached to the result as an attribute \texttt{"stats"}:

\begin{Shaded}
\begin{Highlighting}[]
\CommentTok{\# Below, we use the fact that logit(2/3) = log(2)}
\NormalTok{triangles \textless{}{-}}\StringTok{ }\ControlFlowTok{function}\NormalTok{(nwState, ...) }\KeywordTok{summary}\NormalTok{(}\KeywordTok{as.network}\NormalTok{(nwState) }\OperatorTok{\textasciitilde{}}\StringTok{ }\NormalTok{triangles)}
\NormalTok{nw10 \textless{}{-}}\StringTok{ }\KeywordTok{network}\NormalTok{(}\DecValTok{10}\NormalTok{, }\DataTypeTok{directed =} \OtherTok{FALSE}\NormalTok{)}
\NormalTok{out \textless{}{-}}\StringTok{ }\KeywordTok{simulate}\NormalTok{(nw10 }\OperatorTok{\textasciitilde{}}\StringTok{ }\NormalTok{edges, }\DataTypeTok{nsim =} \DecValTok{50}\NormalTok{, }\DataTypeTok{coef =} \KeywordTok{log}\NormalTok{(}\DecValTok{2}\NormalTok{), }\DataTypeTok{output =}\NormalTok{ triangles)}
\KeywordTok{unlist}\NormalTok{(out, }\DataTypeTok{use.names =} \OtherTok{FALSE}\NormalTok{)}
\end{Highlighting}
\end{Shaded}

\begin{verbatim}
##  [1] 22 78 41 45 22 50 34 42 42 32 37 48 38 28 65 17 23 29 40 23 66 24 26 39 35 34 33 39 35 39
## [31] 61 27 35 23 32 64 34 20 40 56 30 44 34 36 72 42 30 43 11 32
\end{verbatim}

\begin{Shaded}
\begin{Highlighting}[]
\KeywordTok{head}\NormalTok{(}\KeywordTok{attr}\NormalTok{(out, }\StringTok{"stats"}\NormalTok{))}
\end{Highlighting}
\end{Shaded}

\begin{verbatim}
##      edges
## [1,]    26
## [2,]    39
## [3,]    32
## [4,]    31
## [5,]    26
## [6,]    33
\end{verbatim}

\hypertarget{mcmc-proposal-hints}{%
\subsection{MCMC Proposal hints}\label{mcmc-proposal-hints}}

\label{sec:MCMChints}

In Equation \eqref{metropolishastings}, basically any proposal
distribution \(q\) that can, through multiple steps, reach any given
point in \(\netsY\) leads in theory to a Markov chain with the correct
stationary distribution. In practice, however, some choices of \(q\) may
result in rejection of nearly all proposals, a situation informally
called ``slow mixing.'' Thus, it is helpful to have access to different
proposals to deploy in different situations.

For example, most real-world social networks are sparse: The vast
majority of potential ties are not realized. This results in the basic
uniform proposal spending a lot of computing effort proposing toggles to
non-edges that are rejected by the Metropolis--Hastings algorithm.

The TNT, or tie/no tie, proposal was introduced in the \pkg{ergm}
package specifically to address this slow mixing due to sparsity. TNT
proposes a toggle to a uniformly randomly chosen existing edge with
probability 1/2, or a toggle to a uniformly randomly chosen
dyad---including both edges and non-edges---with probability 1/2. (The
description of TNT in Morris et al. (2008) is slightly erroneous; under
TNT, the probability of proposing a non-edge for toggling is less than
1/2, not equal to 1/2.) For sparse networks, the set of edges is much
smaller than the set of non-edges, so TNT speeds mixing by making many
more `off'-toggle proposals than would occur if all dyads had the same
probability of being proposed for a toggle.

\pkg{ergm} 4 introduces a concept of a \emph{hint} to enable the user to
inform the sampling and estimation algorithm about the properties of the
network model that, while they do not affect its stationary
distribution, can be helpful in sampling. This information is specified
via the \texttt{MCMC.prop=} or \texttt{obs.MCMC.prop=} control
parameters as a one-sided formula. For example,
\texttt{MCMC.prop=\textasciitilde{}sparse} (the default) informs the
proposal selection algorithm that the sampling should be optimized for
sparse networks, which typically means enabling the above-described TNT
algorithm. As another example, the code in \secref{sec:Efficiency}
includes the following line within an \texttt{ergm} call:

\begin{Shaded}
\begin{Highlighting}[]
\NormalTok{MCMC.prop =}\StringTok{ }\ErrorTok{\textasciitilde{}}\KeywordTok{strat}\NormalTok{(}\DataTypeTok{attr =} \OperatorTok{\textasciitilde{}}\NormalTok{race, }\DataTypeTok{empirical =} \OtherTok{TRUE}\NormalTok{) }\OperatorTok{+}\StringTok{ }\NormalTok{sparse}
\end{Highlighting}
\end{Shaded}

The \texttt{strat} hint in this case instructs the proposal distribution
to take the \texttt{race} attribute of nodes into account when proposing
dyads to toggle; in particular, \texttt{empirical\ =\ TRUE} instructs
the proposal to weight every possible node-pair \texttt{race}
combination according to the proportions of such combinations observed
in the network used at the beginning of the Markov chain. Alternatively,
the user may pass the \texttt{strat} hint an explicit matrix of weights
via the \texttt{pmat} argument. Additional information about hints
currently implemented in \pkg{ergm} is available via
\texttt{?\ ergmHint}, with \texttt{help("{[}name{]}-ergmHint")} or
\texttt{ergmHint?{[}name{]}} for more details about a specific hint.

\hypertarget{mcmc-proposal-constraints}{%
\subsection{MCMC Proposal constraints}\label{mcmc-proposal-constraints}}

\label{sec:MCMCconstraints}

It is possible to constrain the sample space of possible networks
allowed to have positive probability under the ERGM or, equivalently, to
define a subset of networks as having zero probability under the model.
Technically, such constraints need not influence our choice of \(q\) in
Equation \eqref{metropolishastings}, since any proposed network whose
probability under the model is zero cannot be accepted by the
Metropolis--Hastings algorithm. However, in practice it is a waste of
computing effort to propose such networks in the first place. The
\pkg{ergm} package allows certain types of constraints to be respected
by \(q\), leading to substantial gains in efficiency when these
constraints exist.

The new \texttt{ergm} proposal \texttt{BDStratTNT}, in addition to
supporting the \texttt{strat} and \texttt{sparse} hints described in
\secref{sec:MCMChints}, allows the user to fix edge states of dyads of
specified mixing types according to a vertex attribute via the
\texttt{blocks} constraint. It also allows for upper bounds on a node's
degree via the \texttt{bd} constraint's \texttt{maxout}, \texttt{maxin},
and \texttt{attribs} arguments. Documentation for all MCMC proposals
visible to \pkg{ergm} is available via \texttt{?ergmProposal} and, for a
specific proposal, \texttt{help("{[}name{]}-ergmProposal")} or the
shorthand \texttt{ergmProposal?{[}name{]}}.

In addition, the \pkg{tergm} package includes a generalization of
\texttt{BDStratTNT} that specifically supports dynamic models by
considering a dyad's discordance, i.e., whether the dyad is in the same
state that it was in at the beginning of the time step when the Markov
chain for that step was initialized.

As an example application of the \texttt{BDStratTNT} proposal, the
examples of \secref{sec:Efficiency} contain the following line:

\begin{Shaded}
\begin{Highlighting}[]
\NormalTok{constraints =}\StringTok{ }\ErrorTok{\textasciitilde{}}\KeywordTok{bd}\NormalTok{(}\DataTypeTok{maxout =} \DecValTok{1}\NormalTok{) }\OperatorTok{+}\StringTok{ }\KeywordTok{blocks}\NormalTok{(}\DataTypeTok{attr =} \OperatorTok{\textasciitilde{}}\NormalTok{sex, }\DataTypeTok{levels2 =} \KeywordTok{diag}\NormalTok{(}\OtherTok{TRUE}\NormalTok{, }\DecValTok{2}\NormalTok{))}
\end{Highlighting}
\end{Shaded}

This line ensures that the sample space of possible networks includes
only networks in which no node has a degree of 2 or more---the networks
in these examples are undirected---and in which no changes in dyad
status between nodes of the same \texttt{sex} value are allowed. These
constraints are used to model a network of monogamous heterosexual
relationships.

\hypertarget{adaptive-mcmc-via-effective-sample-size}{%
\subsection{Adaptive MCMC via effective sample
size}\label{adaptive-mcmc-via-effective-sample-size}}

\pkg{ergm} 4 implements adaptive MCMC sampling. Monte-Carlo-based
approximate maximum likelihood estimation, sometimes abbreviated MCMLE,
for an ERGM depends only on the MCMC sample of the sufficient statistic
values and is agnostic to the underlying graphs once the statistics have
been calculated (Hunter \& Handcock, 2006; Krivitsky, 2012; Krivitsky \&
Butts, 2017). Furthermore, while different algorithms approach the
problem in different ways, the estimation ultimately entails matching
the mean of the simulated statistic under \(\curvpar\) to the observed
statistic. Thus, the sampling algorithm can focus on obtaining a
particular \emph{effective sample size} of the multivariate sufficient
statistic.

The user or the estimation algorithm specifies the target effective
sample size, typically via the control parameter
\texttt{MCMC.effectiveSize} or, for estimation,
\texttt{MCMLE.effectiveSize}, as well as the initial MCMC thinning
interval (\texttt{MCMC.interval}) and sample size
(\texttt{MCMC.samplesize}). The algorithm then iterates the following
steps:

\begin{enumerate}
\def\labelenumi{\arabic{enumi}.}
\tightlist
\item
  Run the Markov chain
  \texttt{MCMC.samplesize}\(\times\)\texttt{MCMC.interval} steps forward
  to obtain an initial sample of size \texttt{MCMC.samplesize}.
\item
  If the size of the Markov chain's cumulative sample size exceeds
  \(2\times\)\texttt{MCMC.samplesize}, discard every other draw and
  double \texttt{MCMC.interval} for future runs.
\item
  Identify a candidate ``burn-in'' period by fitting a multivariate
  regression model to the sampled statistics or estimating functions.
  That is, considering an MCMC sample of \(S\) statistics
  \(\genstats(\Y^{(1)}),\dotsc,\genstats(\Y^{(S)})\), we find a
  least-squares fit for \[\dnatpar\t\genstats(\Y^{(s)}) =
  \bm{\beta} _ 0 + \bm{\beta} _ 1 2^{-s/s_0} + \bm{e}^{(s)},\
  s=1,\dotsc,S,\] where \(s_0>0\) is the candidate burn-in,
  \(\bm{\beta} _ 0, \bm{\beta} _ 1\in \reals^{\ncurvpar}\) are
  (nuisance) parameters, and \(\bm{e} ^ {(s)}\in \reals^{\ncurvpar}\)
  are the residuals. In practice, this estimator can be determined
  numerically for a given \(s_0\), so we perform a bisection search over
  the possible \(s_0\). Loosely, \(s_0\) is the number of iterations
  needed to reduce the expected difference between the current value of
  the statistic and its equilibrium value by a factor of 2.
\item
  Evaluate a multivariate extension of the Geweke (1991) convergence
  diagnostic after discarding sample units up to the estimated \(s_0\).
\item
  If nonconvergence is detected, repeat from Step 1, accumulating draws.
\item
  Calculate the effective sample size of the retained draws using the
  method of Vats et al. (2019). If satisfied, return.
\item
  Extrapolate to estimate the additional number of Markov chain steps to
  obtain the target effective sample size given the current ratio of the
  sample size to the effective sample size. Advance the estimated number
  of steps, accumulating draws.
\item
  Continue from Step 2.
\end{enumerate}

\hypertarget{maximum-likelihood-estimation-enhancements}{%
\section{Maximum likelihood estimation
enhancements}\label{maximum-likelihood-estimation-enhancements}}

\label{sec:MLEenhancements}

Frequentist inference for ERGMs calls for finding an estimator given
observed data on a network or networks, along with estimates of the
variability of that estimator that are generally expressed in the form
of standard errors. We consider the gold standard of estimation to be
the maximum likelihood estimator (MLE), while the maximum
pseudo-likelihood estimator (MPLE) is an alternative with certain
advantages and disadvantages relative to the MLE. Calculating estimates
like these along with their standard errors is the core functionality of
the \pkg{ergm} package, and in this section we describe multiple
enhancements to the package as of version 4.

The likelihood function is, by definition, the function of Equation
\eqref{ergm} when that expression is viewed as a function of
\(\curvpar\). Also in the likelihood function, we often replace the
generic \(\y\) by \(\yobs\) when an observed network is at hand. The
natural logarithm of the likelihood function is often denoted
\(\llik(\curvpar)\). The MLE \(\mle\) is the maximizer of
\(\llik(\curvpar)\). Alternatively, the MLE is a zero of the gradient of
the log-likelihood, also known as the score function (Hunter \&
Handcock, 2006, Equation 3.1), i.e., \(\mathbf{U}(\mle) = \0\), where
\begin{equation}\label{eq:score}
\mathbf{U}(\curvpar)\defeq
\grad_{\curvpar}\llik(\curvpar)=\dnatpar\t[\genstats(\yobs) - \ETyheg{\curvpar}\genstats(\Y)].
\end{equation}

The normalizing constant \(\cheg(\curvpar,\netsY)\) of Equation
\eqref{ergm} has the property that
\(\grad_{\curvpar} \log \cheg(\curvpar,\netsY) = \ETyheg{\curvpar}\genstats(\Y)\),
which gives rise to \eqref{eq:score}. Often this constant is
computationally intractable, and a number of approaches (e.g., Snijders,
2002; Hummel et al., 2012) have been proposed for approximating the MLE.
The \pkg{ergm} package defaults to the importance sampling approach of
Hunter \& Handcock (2006): The likelihood ratio is expressed as an
expectation with respect to one of the parameter configurations (that of
the \(\guessind\)th guess, denoted \(\curvpar^\guessind\)), and a
simulation from that configuration is used to maximize this ratio with
respect to \(\curvpar\) to obtain the next guess
\(\curvpar^{\guessind+1}\). In particular, \begin{equation}
\nextguess=\argmax_{\curvpar} \en{\bigg(}{\bigg)}{\innerprod{\en\{\}{\natpar-\natparT{\guessind}}}{\genstats(\yobs)} - \log \meansamp
\exp\en[]{\innerprod{\en\{\}{\natpar-\natparT{\guessind}}}{\genstats(\y^{\guess,\sampind})}}},
\label{eq:mcmle-lr}
\end{equation} where \(\y^{\guess,1}, \ldots, \y^{\guess,\sampsize}\) is
an approximate sample from \(\PTyheg{\guess}\) obtained via MCMC. It is
because an MCMC-generated sample is used to obtain an approximate MLE
that this and related procedures are sometimes called MCMCMLE; the fact
that this approach is a central pillar of \pkg{ergm} underscores how
integral the MCMC algorithms are to nearly all facets of the package.

\hypertarget{sec:obs-step}{%
\subsection{Missing data}\label{sec:obs-step}}

Handcock (2003) observed that for a non-curved family, i.e., where
\(\natpar\equiv\curvpar\), there is no maximizer in Equation
\eqref{eq:mcmle-lr} if \(\genstats(\yobs)\) is not in the convex hull of
the sample \(\genstats(\y^{\guess,\sampind})\). This can be seen because
if \(\genstats(\yobs)\) is outside of the convex hull, then one can
increase the maximand arbitrarily by selecting
\(\curvpar'=\alpha \genstats(\yobs) + \curvpar^\guessind\) with
\(\alpha\to\infty\): the first term in \eqref{eq:mcmle-lr} would grow
faster than any summand or combination of summands of the second.
Conversely, if \(\genstats(\yobs)\) is in the interior of the convex
hull, then for any direction for \(\curvpar'\), some combinations of
summands would grow faster than the first term. Hummel et al. (2012)
therefore proposed to translate \(\genstats(\yobs)\) in the direction of
the centroid of \(\genstats(\y^{\guess,\sampind})\) until it is
sufficiently deep inside the convex hull, making an update that would be
guaranteed to be a unique maximizer in the correct general direction.
Krivitsky (2017) extended this approach to curved ERGMs.

When there are missing data or an observation process, as described in
Section 7 of Krivitsky, Hunter, et al. (2022), the form of the maximand
becomes
\[\log \meansamp \exp\en[]{\innerprod{\en\{\}{\natparp-\natparT{\guessind}}}{\genstats(\y^{\guess,\sampind}|\yobs)}} -
\log \meansamp \exp\en[]{\innerprod{\en\{\}{\natparp-\natparT{\guessind}}}{\genstats(\y^{\guess,\sampind})}},\]
where \(\y^{\guess,\sampind}|\yobs\) are draws from the distribution
\(\DY{\netsY(\yobs)}(\guess)\). This expression can be maximized to
infinity if \emph{any} of \(\genstats(\y^{\guess,\sampind}|\yobs)\) is
outside of the convex hull of \(\genstats(\y^{\guess,\sampind})\), as
that summand could then dominate all others in both summations.

\pkg{ergm} therefore scales all
\(\genstats(\y^{\guess,\sampind}|\yobs)\) toward the centroid of
\(\genstats(\y^{\guess,\sampind})\) until they are all sufficiently deep
in the convex hull. That they are being scaled towards a point
guarantees that such a scaling factor exists, unless the rank of
\(\y^{\guess,\sampind}|\yobs\) is higher than that of
\(\genstats(\y^{\guess,\sampind})\). Version 4 of the \pkg{ergm} package
introduces substantial improvements to the algorithm that determines
whether a point is inside the convex hull of a given set of points. For
instance, the algorithm now returns, after solving a single linear
program, the exact multiplier that scales the point so that it lies on
the boundary of the convex hull; furthermore, \pkg{ergm} now uses the
\pkg{Rglpk} package rather than the \pkg{lpSolveAPI} package when the
former is installed, which according to our tests solves the convex hull
linear program substantially more efficiently. Further details of the
improvements to the convex hull algorithm are in Krivitsky, Kuvelkar, et
al. (2022).

\hypertarget{standard-errors-for-maximum-pseudo-likelihood-estimation}{%
\subsection{Standard errors for maximum pseudo-likelihood
estimation}\label{standard-errors-for-maximum-pseudo-likelihood-estimation}}

\label{sec:MPLE}

To define the maximum pseudo-likelihood estimator (MPLE) for a binary
network, we must first define the pseudo-likelihood function. To this
end, let us consider that Equation \eqref{singletoggle} arises due to
the fact that under the ERGM of Equation \eqref{ergm},
\begin{equation}\label{ConditionalLogOdds}
\log \frac { \Ptyheg(\y\cup\enbc{\pij}) }
{ \Ptyheg(\y\setminus\enbc{\pij}) } =
\log \frac { \Ptyheg(\Yij=1 \,|\, \y_{ij}^c) }
{ \Ptyheg(\Yij=0 \,|\, \y_{ij}^c) } =
\cnmap\t \changeij\genstats(\y),
\end{equation} where \(\y_{ij}^c\) is the entirety of \(\y\)
\emph{except} \(\yij\). If we imagine that all \(\Yij\) are mutually
independent Bernoulli random variables with distributions given by
\eqref{ConditionalLogOdds}---i.e., that
\(\logit\Ptyheg(\Yij=1) = \cnmap\t \changeij\genstats(\y)\)---then
multiplying their individual probability mass functions gives a function
called the \emph{pseudo-likelihood}, whose maximizer is known as the
maximum pseudo-likelihood estimator (MPLE).

As discussed in Hunter et al. (2008), the \texttt{ergm} function uses
logistic regression to obtain MPLEs in non-curved models. The
\texttt{glm} function in \proglang{R} returns not only estimated
logistic regression coefficients (parameters) but also standard errors
for those coefficients. Earlier versions of \pkg{ergm} had simply
reported these standard errors without modification whenever the user
asked for MPLE model output; however, these standard errors are
inaccurate when the edges are not independent. Indeed, a straightforward
Taylor approximation (Schmid \& Hunter, 2021) gives
\begin{equation}\label{eq:sandwich}
\VTyheg{\curvpar} (\mle_\text{MPLE}) \approx [J(\curvpar)]^{-1} \VTyheg{\curvpar}({\mathbf U}(\curvpar)) [J(\curvpar)]^{-1},
\end{equation} where we assume that the ERGM with parameter \(\curvpar\)
is the true model for the distribution that gave rise to the observed
network and \(J(\curvpar)\) is the negative Hessian of the logarithm of
the pseudo-likelihood function. Since \(\curvpar\) is of course unknown,
in practice we may approximate the right hand side above by substituting
\(\mle_\text{MPLE}\) for \(\curvpar\).

Whenever the form of \(\genstats(\y)\) results in all \(\Yij\) being
mutually independent, as discussed in Krivitsky, Hunter, et al. (2022),
the MPLE equals the MLE. This also implies
\(J(\curvpar)= \VTyheg{\curvpar}[{\mathbf U}(\curvpar)]\) for all
\(\curvpar\), so the expression on the right hand side simplifies to
\([J(\curvpar)]^{-1}\). Indeed, the standard errors returned by the
logistic regression algorithm are those given by
\([J(\mle_\text{MPLE})]^{-1}\). Yet for non-dyad-independent models,
these logistic-regression-based standard errors are poor approximations
to the true standard deviations of the parameter estimates. The
\pkg{ergm} package therefore implements the technique described by
Schmid \& Hunter (2021), using \eqref{eq:sandwich} with
\(\mle_\text{MPLE}\) in place of \(\curvpar\) to provide standard
errors. For this purpose, the middle term must be estimated by the
sample covariance matrix from a random sample obtained using Markov
chain Monte Carlo with \(\mle_\text{MPLE}\) as the true parameter value.
Alternatively, \texttt{ergm} can approximate MPLE standard errors via a
bootstrap method, as proposed by Schmid \& Desmarais (2017).

Logistic regression to obtain the MPLE is performed automatically when
\texttt{estimate="MPLE"} is used with the \texttt{ergm} function. The
\pkg{ergm} package also provides a function \texttt{ergmMPLE} that
produces the response vector and predictor matrix that may be used, for
instance, to produce the logistic regression output directly via the
\texttt{glm} function in \proglang{R}. The \texttt{ergmMPLE} function
gives its output in the form of weighted response/predictor
combinations, weighted according to their multiplicity, in order to
conserve memory in cases where particular combinations occur frequently.

\begin{Shaded}
\begin{Highlighting}[]
\KeywordTok{data}\NormalTok{(g4)}
\KeywordTok{print}\NormalTok{(lr \textless{}{-}}\StringTok{ }\KeywordTok{ergmMPLE}\NormalTok{(g4 }\OperatorTok{\textasciitilde{}}\StringTok{ }\NormalTok{edges }\OperatorTok{+}\StringTok{ }\NormalTok{triangle))}
\end{Highlighting}
\end{Shaded}

\begin{verbatim}
## $response
## [1] 0 0 0 1 1 1
## 
## $predictor
##      edges triangle
## [1,]     1        1
## [2,]     1        0
## [3,]     1        2
## [4,]     1        0
## [5,]     1        1
## [6,]     1        2
## 
## $weights
## [1] 2 1 4 1 2 2
\end{verbatim}

The weights vector sums to \(4 \times 3\), the number of potential
relations in the network. We may verify that the MPLE obtained via
\texttt{ergm} matches direct logistic regression estimates:

\begin{Shaded}
\begin{Highlighting}[]
\KeywordTok{rbind}\NormalTok{(}\KeywordTok{coef}\NormalTok{(}\KeywordTok{ergm}\NormalTok{(g4}\OperatorTok{\textasciitilde{}}\NormalTok{edges }\OperatorTok{+}\StringTok{ }\NormalTok{triangle, }\DataTypeTok{estimate=}\StringTok{"MPLE"}\NormalTok{)),}
  \KeywordTok{coef}\NormalTok{(}\KeywordTok{glm}\NormalTok{(response }\OperatorTok{\textasciitilde{}}\StringTok{ }\NormalTok{predictor }\OperatorTok{{-}}\StringTok{ }\DecValTok{1}\NormalTok{, }\DataTypeTok{weights  =}\NormalTok{ weights, }\DataTypeTok{data =}\NormalTok{ lr, }\DataTypeTok{family =} \StringTok{"binomial"}\NormalTok{)))}
\end{Highlighting}
\end{Shaded}

\begin{verbatim}
##          edges   triangle
## [1,] 0.2057346 -0.4114692
## [2,] 0.2057346 -0.4114692
\end{verbatim}

The predictor vector associated with dyad \((i,j)\) is the change
statistic vector for that dyad under the model, and the change
statistics can be queried directly to produce an array of change
statistics corresponding to each tail, head, and statistic combination
using the \texttt{output="array"} argument:

\begin{Shaded}
\begin{Highlighting}[]
\KeywordTok{ergmMPLE}\NormalTok{(g4 }\OperatorTok{\textasciitilde{}}\StringTok{ }\NormalTok{edges }\OperatorTok{+}\StringTok{ }\NormalTok{triangle, }\DataTypeTok{output =} \StringTok{"array"}\NormalTok{)}\OperatorTok{$}\NormalTok{predictor}
\end{Highlighting}
\end{Shaded}

\begin{verbatim}
## , , term = edges
## 
##     head
## tail V1 V2 V3 V4
##   V1 NA  1  1  1
##   V2  1 NA  1  1
##   V3  1  1 NA  1
##   V4  1  1  1 NA
## 
## , , term = triangle
## 
##     head
## tail V1 V2 V3 V4
##   V1 NA  0  1  2
##   V2  0 NA  2  1
##   V3  1  2 NA  2
##   V4  2  1  2 NA
\end{verbatim}

Alternatively \texttt{output="dyadlist"} produces an uncompressed list:

\begin{Shaded}
\begin{Highlighting}[]
\KeywordTok{ergmMPLE}\NormalTok{(g4 }\OperatorTok{\textasciitilde{}}\StringTok{ }\NormalTok{edges }\OperatorTok{+}\StringTok{ }\NormalTok{triangle, }\DataTypeTok{output =} \StringTok{"dyadlist"}\NormalTok{)}\OperatorTok{$}\NormalTok{predictor}
\end{Highlighting}
\end{Shaded}

\begin{verbatim}
##       tail head edges triangle
##  [1,]    1    2     1        0
##  [2,]    1    3     1        1
##  [3,]    1    4     1        2
##  [4,]    2    1     1        0
##  [5,]    2    3     1        2
##  [6,]    2    4     1        1
##  [7,]    3    1     1        1
##  [8,]    3    2     1        2
##  [9,]    3    4     1        2
## [10,]    4    1     1        2
## [11,]    4    2     1        1
## [12,]    4    3     1        2
\end{verbatim}

\hypertarget{log-likelihood-estimation}{%
\subsection{Log-likelihood estimation}\label{log-likelihood-estimation}}

\label{sec:LogLikelihood}

Likelihood-based estimation relies not only on the value of the
maximizer \(\mle\) of the log-likelihood function, but also on the
maximum value \(\llik(\mle)\) that function obtains. Model selection
criteria such as AIC and BIC are based on this maximized
log-likelihood---they are equal to \(-2\llik(\mle) + 2p\) and
\(-2\llik(\mle) + p\log d\), respectively, where \(p\) is the number of
model parameters and \(d\) is the number of observed, non-fixed
potential relations in the network---as are the standard chi-squared
tests based on drop-in-deviance.

Hunter \& Handcock (2006) point out that the null deviance, which is
equal to \(-2\llik(\0)\), is straightforward to calculate; in the case
of binary networks, it equals \(2N\log 2\), where again \(N\) is the
number of observed, non-fixed potential edges. Yet log-likelihood values
are sometimes computationally intractable, as mentioned in
\secref{sec:Introduction}. A novel method currently employed by the
\pkg{ergm} package is to first identify all dyadic dependent terms in
the model, then find the MLE and corresponding log-likelihood value in
the constrained parameter space that fixes the values of the
coefficients corresponding to those terms at zero. This calculation is
straightforward using logistic regression, as explained in
\secref{sec:MPLE}. If we denote the MLE of this sub-model as \(\mple\),
then our task becomes estimation of \(\ell(\mle)-\ell(\mple)\), since
the second term in this expression is known from logistic regression
output.

Section 5 of Hunter \& Handcock (2006) addresses the problem of
likelihood ratio testing, which on the logarithmic scale is exactly the
problem of calculating the difference of two log-likelihoods such as
\(\ell(\mle)-\ell(\mple)\). That paper describes the idea of path
sampling (Gelman \& Meng, 1998), which is based on the following
observation: if we define a smooth path in parameter space from
\(\mple\) to \(\mle\), that is, a differentiable function \(\mathbf{m}\)
that maps the closed unit interval \([0,1]\) into the parameter space so
that \(\mathbf{m}(0)=\mple\) and \(\mathbf{m}(1)=\mle\), then
\begin{equation}
\label{PathSampling}
\log\cheg(\mle,\netsY)-\log\cheg(\mple,\netsY) =
\int_0^1 \ETyheg{\mathbf{m}(u)}\ENBC{{\frac{d}{du}} \cnmap[\mathbf{m}(u)]}\t\genstats(\Y) \,du
\end{equation} by the fundamental theorem of calculus, where
\(\cheg(\curvpar,\netsY)\) is the normalizing constant of Equation
\eqref{ergm}. Pulling the expectation and differentiation operators
outside of the integral, the expression remaining under the integral
sign is also an expectation with respect to a random variable \(U\)
uniformly distributed on \((0,1)\). Thus, Equation \eqref{PathSampling}
implies that \begin{equation}
\label{PathSampling2}
\log\cheg(\mle,\netsY)-\log\cheg(\mple,\netsY) =
\E_{\netsY,\h,\cnmap,\genstats} \ENBC{\frac{d}{dU} \cnmap[\mathbf{m}(U)]}\t\genstats(\Y),
\end{equation} where the expectation is taken with respect to the joint
distribution of \(U\) and \(\Y\), where \(U\sim\Uniform(0,1)\) and
\(\Y \given U\) is distributed according to the ERGM of Equation
\eqref{ergm} with parameter \(\mathbf{m}(U)\). Here, we introduce a
shifted version of the vector \(\genstats(\cdot)\) of sufficient
statistics, \[
\diffstats(\y) \defeq \genstats(\y) - \genstats(\yobs),
\] where if missing data are present we replace \(\genstats(\yobs)\) by
\(\E_{\curvpar; \netsY,\h,\cnmap,\genstats}\genstats(\Y \given \yobs)\);
yet here we assume for simplicity of notation that \(\diffstats(\y)\)
does not depend on \(\curvpar\). If \(\diffstats(\cdot)\) is substituted
for \(\genstats(\cdot)\), then
\(\ell(\curvpar) = -\log\normc_{\h,\cnmap,\diffstats}(\curvpar, \netsY)\),
so for instance Equation \eqref{PathSampling2} gives a convenient
expression for \(\ell(\mple) - \ell(\mle)\).

In practice, the problem with Equation \eqref{PathSampling2} is that
simulating the first network \(\Y\) from a given parameter configuration
\(\curvpar=\mathbf{m}(U)\) requires a long ``burn-in'' period, which
makes the direct approach of drawing a \(U_k\) then a \(\Y_k|U_k\) for
\(k=1,\dotsc,K\) impractical. On the other hand, once ``burned in,''
subsequent draws \(\Y_2, \dotsc, \Y_K\) from the same distribution cost
relatively little additional effort. For this reason, the \pkg{ergm}
package currently implements a technique known as bridge sampling (Meng
\& Wong, 1996) as an approximation of Equation \eqref{PathSampling2}.

Bridge sampling in \pkg{ergm} partitions the unit interval into \(J\)
sub-intervals, each of length \(1/J\), where \(u_j\) is taken to be the
center of the \(j\)th sub-interval for \(j = 1,\dotsc,J\). For each
\(j\), we simulate a random sample \(\Y_{j1}, \ldots, \Y_{jK}\) of
networks from the ERGM with parameter \(\mathbf{m}(u_j)\). Since the
\(u_j\) values may be viewed as a rough approximation of a uniform
sample on \([0,1]\), the idea of Equation \eqref{PathSampling2} leads to
\begin{equation}
\label{BridgeSampling}
\ell(\mle)-\ell(\mple) \approx
- \frac{1}{JK} \sum_{j=1}^J \sum_{k=1}^K
\nabla \mathbf{m}(u_j) \nabla \cnmap [\mathbf{m}(u_j)] \diffstats(\Y_{jk}).
\end{equation} (The analogous Equation 5.4 of Hunter \& Handcock (2006)
omits the needed factor \(-1/J\).)

Equation \eqref{BridgeSampling} entails two different approximations of
Equation \ref{PathSampling2}: One in approximating the expectation of
\(\Y\) using simulated networks and one in approximating the
\(\Uniform(0,1)\) distribution by \(u_0, \ldots, u_J\). The first of
these is due to Monte Carlo error, so it may be quantified; this is the
source of the standard errors reported for AIC and BIC for
dyadic-dependent models. The user can specify a control parameter
\texttt{bridge.target.se}, via \texttt{control.logLik.ergm()} or
\texttt{snctrl()}, to continue bridge sampling at values \(u_{l,j}\) for
\(j=1,\dotsc,J\) and \(l=1,2,\dotsc\) \emph{ad infinitum}, until the
estimated standard error due to the first approximation is below it.

The second approximation leads to a bias even in the idealized case
where \(K\to\infty\), which only vanishes as \(J\to\infty\). Our
experiments suggest that this bias is small, but in the adaptive mode
triggered by \texttt{bridge.target.se}, it is addressed by shifting each
series of \(J\) points by a value from a low-discrepancy sequence in one
dimension: \(u _ {l,j} = (j - 1/2 + v _ l)/J\), for
\(v _ l = \modop((l-1)/\phi + 1/2, 1) - 1/2\), a shifted Kronecker
sequence with \(v _ 1=0\) and the inverse of the Golden Ratio \(\phi\)
as its coefficient. The ``weight'' of each \(u _ {l,j}\) in Equation
\ref{BridgeSampling} is then adjusted to be proportional to the size of
its Voronoi cell, which is the length of the region of points on the
unit interval that are nearer to \(u _ {l,j}\) than to any other point
\(u _ {l',j'}\) sampled so far.

As a further optimization, the algorithm reduces the need for burning-in
by reordering points \(u _ {l,j}\) for a given \(l\) such that
\(u _ {l+1,1}\) is as close to \(u_{l,J}\) as possible, \(u_{l+1,2}\) to
\(u_{l+1,1}\), and so on.

\hypertarget{curved-mple-and-curved-ergms-as-first-class-models}{%
\subsection{Curved MPLE and curved ERGMs as ``first-class''
models}\label{curved-mple-and-curved-ergms-as-first-class-models}}

\label{sec:CurvedMPLE}

Curved ERGMs---those for which \(\natpar\ne\curvpar\)---were introduced
by Hunter \& Handcock (2006) to facilitate estimation of the decay
parameter in the geometrically-weighted triadic degree and triadic
terms. They stated a score function for such models and outlined an
MCMLE algorithm that could be used to update \(\curvpar\) given a sample
of sufficient statistics from the previous guess.

However, the implementation prior to \pkg{ergm} 4 was incomplete in the
following respect: While it could estimate curved ERGMs, it required the
end-user to specify the initial values for their parameters. This is
because maximum pseudo-likelihood estimation (MPLE) for curved models
had not been derived and implemented. Non-MCMLE methods did not support
curved ERGMs at all.

The score function for the curved ERGM MPLE was derived by Krivitsky
(2017) and is implemented in \pkg{ergm} 4. Thus, to fit a curved model
with geometrically weighted degree, one previously had to specify an
initial value, as in:

\begin{Shaded}
\begin{Highlighting}[]
\KeywordTok{data}\NormalTok{(florentine)}
\KeywordTok{ergm}\NormalTok{(flomarriage }\OperatorTok{\textasciitilde{}}\StringTok{ }\NormalTok{edges }\OperatorTok{+}\StringTok{ }\KeywordTok{gwdegree}\NormalTok{(}\FloatTok{0.25}\NormalTok{))  }\CommentTok{\# Initial guess for the decay parameter = 0.25}
\end{Highlighting}
\end{Shaded}

Now, the initial value is determined automatically if we simply specify

\begin{Shaded}
\begin{Highlighting}[]
\KeywordTok{data}\NormalTok{(florentine)}
\KeywordTok{ergm}\NormalTok{(flomarriage }\OperatorTok{\textasciitilde{}}\StringTok{ }\NormalTok{edges }\OperatorTok{+}\StringTok{ }\NormalTok{gwdegree)}
\end{Highlighting}
\end{Shaded}

\begin{verbatim}
## 
## Call:
## ergm(formula = flomarriage ~ edges + gwdegree)
## 
## Last MCMC sample of size 486 based on:
##          edges        gwdegree  gwdegree.decay  
##        -1.5334         -0.1318          0.6730  
## 
## Monte Carlo Maximum Likelihood Coefficients:
##          edges        gwdegree  gwdegree.decay  
##     -1.6255705      -0.0008501       0.2259348
\end{verbatim}

In situations where the decay parameter is fixed and known, its value
may be specified directly or via the partial \texttt{offset} capability,
also new in \pkg{ergm} 4. Thus, the two \texttt{ergm} calls below are
equivalent, though their coefficient estimates differ slightly because
the fitting algorithm is stochastic:

\begin{Shaded}
\begin{Highlighting}[]
\KeywordTok{data}\NormalTok{(sampson)}
\KeywordTok{coef}\NormalTok{(}\KeywordTok{ergm}\NormalTok{(samplike}\OperatorTok{\textasciitilde{}}\NormalTok{edges}\OperatorTok{+}\KeywordTok{gwesp}\NormalTok{(}\FloatTok{0.25}\NormalTok{, }\DataTypeTok{fix=}\OtherTok{TRUE}\NormalTok{), }\DataTypeTok{control=}\KeywordTok{snctrl}\NormalTok{(}\DataTypeTok{MCMLE.maxit=}\DecValTok{2}\NormalTok{)))}
\end{Highlighting}
\end{Shaded}

\begin{verbatim}
##            edges gwesp.fixed.0.25 
##       -1.6374758        0.4136882
\end{verbatim}

\begin{Shaded}
\begin{Highlighting}[]
\KeywordTok{coef}\NormalTok{(}\KeywordTok{ergm}\NormalTok{(samplike}\OperatorTok{\textasciitilde{}}\NormalTok{edges}\OperatorTok{+}\KeywordTok{offset}\NormalTok{(}\KeywordTok{gwesp}\NormalTok{(), }\KeywordTok{c}\NormalTok{(}\OtherTok{FALSE}\NormalTok{,}\OtherTok{TRUE}\NormalTok{)), }\DataTypeTok{offset.coef=}\FloatTok{0.25}\NormalTok{,}
          \DataTypeTok{control=}\KeywordTok{snctrl}\NormalTok{(}\DataTypeTok{MCMLE.maxit=}\DecValTok{2}\NormalTok{)))}
\end{Highlighting}
\end{Shaded}

\begin{verbatim}
##               edges               gwesp offset(gwesp.decay) 
##          -1.6152582           0.4011794           0.2500000
\end{verbatim}

\hypertarget{contrastive-divergence}{%
\subsection{Contrastive Divergence}\label{contrastive-divergence}}

\label{sec:CD}

Contrastive divergence (CD) is a technique taken from the computer
science literature and proposed in the context of ERGMs by Asuncion et
al. (2010). Much more detail about its use for ERGMs is provided by
Krivitsky (2017). Essentially, CD provides a spectrum of estimation
algorithms with MPLE at one extreme and MLE at the other. Little is
presently known about the efficacy of algorithms lying between these two
extremes.

The \pkg{ergm} package implements CD estimates, which may be obtained by
passing \texttt{estimate="CD"} to the \texttt{ergm} function. Since not
much is known about the quality of these estimates, their most promising
use at present is as starting values for MCMC-based maximum likelihood
estimation as described at the beginning of
\secref{sec:MLEenhancements}; that is, they are used for the initial
guess \(\guess\) for \(t=0\). By default, they are used where available
implementations of MPLE are inapplicable, in particular valued ERGMs or
binary ERGMs with dyad-dependent sample space constraints: unlike MPLE,
which must be rederived for each reference distribution and sample space
constraint, contrastive divergence can reuse the proposals from the MCMC
implementation (Krivitsky, 2017).

\hypertarget{confidence-stopping-criterion}{%
\subsection{Confidence stopping
criterion}\label{confidence-stopping-criterion}}

\label{sec:StoppingCriterion}

The \pkg{ergm} package implements several methods to determine when to
declare convergence in an MCMLE algorithm and report the results. They
are selected via the \texttt{MCMLE.termination} parameter as follows:

\begin{description}
\item[\texttt{MCMLE.termination="Hotelling"}]
Convergence is declared if an autocorrelation-adjusted Hotelling \(T^2\)
test is unable to reject the null hypothesis that the estimating
function equals zero, which for non-curved models on fully observed
networks means simply that the expected value of the simulated statistic
equals the observed statistic at a high level (\(\alpha=0.5\) by
default). Krivitsky (2017) provides additional details.
\item[\texttt{MCMLE.termination="Hummel"}]
The algorithm of Hummel et al. (2012) is used: Convergence is declared
if for two consecutive parameter updates, the observed statistic is
sufficiently deep in the interior of the convex hull of the sample of
simulated statistics. (For curved models, sample values of estimating
functions are used instead.) However, this criterion can be problematic
for partially observed networks, as it requires that \emph{every} point
in the constrained (conditional) sample be in the interior of the convex
hull of the unconstrained sample, which can be problematic when the
fraction of the missing dyads and the dimension of the parameter vector
are moderately high.
\item[\texttt{MCMLE.termination="confidence"}]
Loosely based on the algorithms of Vats et al. (2019), this method,
which is the default in \pkg{ergm} 4, implements a form of equivalence
testing. The general idea of equivalence testing is to define the null
hypothesis to be that the difference between the observed and the
expected statistics large enough to be interesting. Thus, rejecting the
null hypothesis entails deciding that the difference is small enough
that convergence is declared.
\end{description}

\hypertarget{simulated-annealing}{%
\section{Simulated Annealing}\label{simulated-annealing}}

\label{sec:SimulatedAnnealing}

\pkg{ergm} has enhanced its flexibility in the use of simulated
annealing (SAN) to randomly generate networks with a particular set of
network statistics. This capability is used by the package in the
process of finding an MLE, particularly when estimating from sufficient
statistics alone rather than an entire network. At least some of the
methods enabled by SAN are the subject of ongoing research. For
instance, Schmid \& Hunter (2020) find that SAN can be used to find
effective starting values for the iterative algorithm described at the
beginning of \secref{sec:MLEenhancements}. The rest of this section
describes simulated annealing and details some of its capabilities and
uses within the \pkg{ergm} package.

\hypertarget{formulation-of-san-algorithm}{%
\subsection{Formulation of SAN
algorithm}\label{formulation-of-san-algorithm}}

\label{sec:SANalgorithm}

Let \(\genstats\) be a vector of target statistics for the network we
wish to construct. That is, we are given an arbitrary network
\(\y^0\in\netsY\), and we seek a network \(\y\in\netsY\) such that
\(\genstats(\y)\approx\genstats\)---ideally equality is achieved, but in
practice we may have to settle for a close approximation. The variant of
simulated annealing used in \pkg{ergm} is as follows.

The energy function is defined
\(\energy_W(\y)=\en\{\}{\genstats(\y)-\genstats}\t W\en\{\}{\genstats(\y)-\genstats}\),
with \(W\) a symmetric positive (barring multicollinearity in
statistics) definite matrix of weights. This function achieves 0 only if
the target is reached. A good choice of this matrix yields a more
efficient search.

A standard simulated annealing loop is used, as described below, with
some modifications. In particular, we allow the user to specify a vector
of offsets \(\cnmap\) to bias the annealing, with \(\cnmapel_k=0\)
denoting no offset. As illustrated in the example below, offsets can be
used with SAN to forbid certain statistics from ever increasing or
decreasing. As with \texttt{ergm}, offset terms are specified using the
\texttt{offset()} decorator and their coefficients specified with the
\texttt{offset.coef} argument. By default, finite offsets are ignored
by, but this can be overridden by setting the control argument
\texttt{SAN.ignore.finite.offsets\ =\ FALSE}.

The number of simulated annealing runs is specified by the
\texttt{SAN.maxit} control parameter and the initial value of the
temperature \(T\) is set to \texttt{SAN.tau}. The value of \(T\)
decreases linearly until \(T=0\) at the last run, which implies that all
proposals that increase \(\energy_W(\y)\) are rejected. The weight
matrix \(W\) is initially set to \(I_{\nnatpar}/\nnatpar\), where
\(I_{\nnatpar}\) is the identity matrix of an appropriate dimension.

For weight \(W\) and temperature \(T\), the simulated annealing
iteration proceeds as follows:

\begin{enumerate}
\def\labelenumi{\arabic{enumi}.}
\tightlist
\item
  Test if \(\energy_W(\y)=0\). If so, then exit.
\item
  Generate a perturbed network \(\y^\star\) from a proposal that
  respects the model constraints. (This is typically the same proposal
  as that used for MCMC.)
\item
  Store the quantity \(\genstats(\y^\star)-\genstats(\y)\) for later
  use.
\item
  Calculate acceptance probability
  \[\alpha = \exp\en[]{-\en\{\}{\energy_W(\y^\star)-\energy_W(\y)}/T + \innerprod{\cnmap}{\en\{\}{\genstats(\y^\star)-\genstats(\y)}}}.\]
  (If \(\abs{\cnmapel_k}=\infty\) and
  \(\genstatel_k(\y^\star)-\genstatel_k(\y)=0\), their product is
  defined to be \(0\).)
\item
  Replace \(\y\) with \(\y^\star\) with probability \(\min(1,\alpha)\).
\end{enumerate}

After the specified number of iterations, \(T\) is updated as described
above, and \(W\) is recalculated by first computing a matrix \(S\), the
sample covariance matrix of the proposed differences stored in Step 3
(i.e., whether or not they were rejected), then \(W=S^+ / \tr(S^+)\),
where \(S^+\) is the Moore--Penrose pseudoinverse of \(S\). The
differences in Step 3 closely reflect the relative variances and
correlations among the network statistics.

In Step 2, the many options for MCMC proposals, including those newly
added to the \pkg{ergm} package as covered in \secref{sec:NewProposals},
can provide for effective means of speeding the SAN algorithm's search
for a viable network. This phenomenon is illustrated in
\secref{sec:SANspeedup}.

The example below illustrates the use of offsets in a 100-node network
in which each node has a \texttt{sex} attribute with possible values
\texttt{"M"} and \texttt{"F"}. Suppose that we wish to construct a
network with 30 edges in which no edges are allowed between nodes of the
same \texttt{sex} value, nor that result in any node having more than
one edge---constraints that would arise, for example, if we wished to
model a network of heterosexual, monogamous relationships. SAN can find
such a network by placing offset parameters valued at \texttt{-Inf} on
ERGM terms corresponding to \texttt{nodematch("sex")} and
\texttt{concurrent}:

\begin{Shaded}
\begin{Highlighting}[]
\NormalTok{nw \textless{}{-}}\StringTok{ }\KeywordTok{network.initialize}\NormalTok{(}\DecValTok{100}\NormalTok{, }\DataTypeTok{directed =} \OtherTok{FALSE}\NormalTok{)}
\NormalTok{nw }\OperatorTok{\%v\%}\StringTok{ "sex"}\NormalTok{ \textless{}{-}}\StringTok{ }\KeywordTok{rep}\NormalTok{(}\KeywordTok{c}\NormalTok{(}\StringTok{"M"}\NormalTok{,}\StringTok{"F"}\NormalTok{), }\DecValTok{50}\NormalTok{)}
\NormalTok{example \textless{}{-}}\StringTok{ }\KeywordTok{san}\NormalTok{(nw }\OperatorTok{\textasciitilde{}}\StringTok{ }\NormalTok{edges }\OperatorTok{+}\StringTok{ }\KeywordTok{offset}\NormalTok{(}\KeywordTok{nodematch}\NormalTok{(}\StringTok{"sex"}\NormalTok{)) }\OperatorTok{+}\StringTok{ }\KeywordTok{offset}\NormalTok{(concurrent),}
               \DataTypeTok{offset.coef =} \KeywordTok{c}\NormalTok{(}\OperatorTok{{-}}\OtherTok{Inf}\NormalTok{, }\OperatorTok{{-}}\OtherTok{Inf}\NormalTok{), }\DataTypeTok{target.stats =} \DecValTok{30}\NormalTok{)}
\KeywordTok{summary}\NormalTok{(example }\OperatorTok{\textasciitilde{}}\StringTok{ }\NormalTok{edges }\OperatorTok{+}\StringTok{ }\KeywordTok{nodematch}\NormalTok{(}\StringTok{"sex"}\NormalTok{) }\OperatorTok{+}\StringTok{ }\NormalTok{concurrent)}
\end{Highlighting}
\end{Shaded}

\begin{verbatim}
##         edges nodematch.sex    concurrent 
##            30             0             0
\end{verbatim}

The output of the \texttt{summary} function above verifies that the
constraints are satisfied by the generated network \texttt{example}.

\hypertarget{computing-efficiency-tests-on-large-networks}{%
\section{Computing efficiency tests on large
networks}\label{computing-efficiency-tests-on-large-networks}}

\label{sec:Efficiency}

Version 4 of the \pkg{ergm} package enables substantial gains in
computing efficiency relative to earlier versions of the packages. There
are many reasons for these gains, including better algorithms (e.g.,
improvements to simulated annealing and maximum pseudo-likelihood
estimation), better use of parallelism, and new MCMC proposals. As
pointed out elsewhere, improved MCMC proposals lead to performance
improvements across a wide range of \pkg{ergm} package functionality due
to the pervasive use of simulation in the \pkg{ergm} workflow. Where
these improvements have greatest impact, we see speedups of as much as
two orders of magnitude for comparable computing tasks. This section
highlights some of the key changes and demonstrates their impacts, using
in some cases networks with one million nodes.

The code used to produce the results in each subsection of
\secref{sec:Efficiency} is given in that subsection. Each test is based
on a hypothetical network constructed from the \texttt{cohab} dataset in
the \pkg{ergm} package, where the nodes are persons and the edges
represent cohabiting pairs. The distribution of nodal attributes and
edges is based on aggregated statistics from the National Survey of
Family Growth (NSFG) (National Center for Health Statistics, 2020). The
nodes have demographic attributes---sex, age, and
race/ethnicity/immigration status---sampled using the NSFG
post-stratification weights that have been adjusted to match the
demographics of King County in Washington State. The edges represent
heterosexual cohabitation relationships observed in the data, which
imposes two constraints on the network that we can exploit for computing
efficiency gains: the networks are bipartite, i.e., only edges between
male and female nodes are allowed, and nodal degree is capped at one. In
addition to demographic attributes, the test dataset includes two more
nodal variables as distributed in the adjusted NSFG data: sexual
identity and whether the node has at least one persistent non-cohabiting
partner. The data and model are a simplified version of an actual
applied research project that models the spread of HIV. Additional
information about these data are available by typing
\texttt{help(cohab)}.

\hypertarget{impact-of-new-proposals-on-markov-chain-mixing}{%
\subsection{Impact of new proposals on Markov chain
mixing}\label{impact-of-new-proposals-on-markov-chain-mixing}}

\label{sec:NewProposals}

This section examines the evolution of a Markov chain based on a
particular ERGM for a million-node network whose nodal covariates are
based on the demographic statistics of the \texttt{cohab} dataset. The
Markov chain is initialized at an empty network, i.e., one with no
edges, and we compare the approach of the ERGM statistics to a set of
target values using each of three different proposals. Again, these
target values are based on the \texttt{cohab} dataset. The model has 15
statistics, defined by the formula used for the simulation:

\begin{Shaded}
\begin{Highlighting}[]
\KeywordTok{data}\NormalTok{(cohab)}
\NormalTok{CohabFormula \textless{}{-}}
\StringTok{  }\NormalTok{nw }\OperatorTok{\textasciitilde{}}\StringTok{ }\NormalTok{edges }\OperatorTok{+}\StringTok{ }\KeywordTok{nodefactor}\NormalTok{(}\StringTok{"sex.ident"}\NormalTok{, }\DataTypeTok{levels =} \DecValTok{3}\NormalTok{) }\OperatorTok{+}\StringTok{ }\KeywordTok{nodecov}\NormalTok{(}\StringTok{"age"}\NormalTok{) }\OperatorTok{+}\StringTok{ }\KeywordTok{nodecov}\NormalTok{(}\StringTok{"agesq"}\NormalTok{) }\OperatorTok{+}
\StringTok{       }\KeywordTok{nodefactor}\NormalTok{(}\StringTok{"race"}\NormalTok{, }\DataTypeTok{levels =} \DecValTok{{-}5}\NormalTok{) }\OperatorTok{+}\StringTok{ }\KeywordTok{nodefactor}\NormalTok{(}\StringTok{"othr.net.deg"}\NormalTok{, }\DataTypeTok{levels =} \DecValTok{{-}1}\NormalTok{) }\OperatorTok{+}
\StringTok{       }\KeywordTok{nodematch}\NormalTok{(}\StringTok{"race"}\NormalTok{, }\DataTypeTok{diff =} \OtherTok{TRUE}\NormalTok{) }\OperatorTok{+}\StringTok{ }\KeywordTok{absdiff}\NormalTok{(}\StringTok{"sqrt.age.adj"}\NormalTok{)}
\end{Highlighting}
\end{Shaded}

Additionally, as explained above, constraints are imposed to prevent
edges between nodes with the same \texttt{sex} attribute and also to
prevent concurrent partnerships, so a node's degree in this network can
be only 0 or 1.

The code below prepares the simulation by estimating ERGM parameters for
a network of 50,000 nodes that give mean statistics equal to the target
statistics based on data collected in King County.

\begin{Shaded}
\begin{Highlighting}[]
\NormalTok{net\_size \textless{}{-}}\StringTok{ }\DecValTok{50000}
\KeywordTok{set.seed}\NormalTok{(}\DecValTok{0}\NormalTok{)}
\NormalTok{inds \textless{}{-}}\StringTok{ }\KeywordTok{sample}\NormalTok{(}\KeywordTok{seq\_len}\NormalTok{(}\KeywordTok{NROW}\NormalTok{(cohab\_PopWts)), net\_size, }\OtherTok{TRUE}\NormalTok{, cohab\_PopWts}\OperatorTok{$}\NormalTok{weight)}
\ControlFlowTok{if}\NormalTok{(RunMode }\OperatorTok{!=}\StringTok{ "Skip"}\NormalTok{) \{}
\NormalTok{  nw \textless{}{-}}\StringTok{ }\KeywordTok{network.initialize}\NormalTok{(net\_size, }\DataTypeTok{directed =} \OtherTok{FALSE}\NormalTok{)}
  \KeywordTok{set.vertex.attribute}\NormalTok{(nw, }\KeywordTok{names}\NormalTok{(cohab\_PopWts)[}\OperatorTok{{-}}\DecValTok{1}\NormalTok{], cohab\_PopWts[inds,}\OperatorTok{{-}}\DecValTok{1}\NormalTok{])}
\NormalTok{  fit \textless{}{-}}\StringTok{ }\KeywordTok{ergm}\NormalTok{(nw }\OperatorTok{\textasciitilde{}}\StringTok{ }\NormalTok{edges }\OperatorTok{+}\StringTok{ }\KeywordTok{nodefactor}\NormalTok{(}\StringTok{"sex.ident"}\NormalTok{, }\DataTypeTok{levels =} \DecValTok{3}\NormalTok{) }\OperatorTok{+}\StringTok{ }\KeywordTok{nodecov}\NormalTok{(}\StringTok{"age"}\NormalTok{) }\OperatorTok{+}\StringTok{ }\KeywordTok{nodecov}\NormalTok{(}\StringTok{"agesq"}\NormalTok{) }\OperatorTok{+}
\StringTok{                   }\KeywordTok{nodefactor}\NormalTok{(}\StringTok{"race"}\NormalTok{, }\DataTypeTok{levels =} \DecValTok{{-}5}\NormalTok{) }\OperatorTok{+}\StringTok{ }\KeywordTok{nodefactor}\NormalTok{(}\StringTok{"othr.net.deg"}\NormalTok{, }\DataTypeTok{levels =} \DecValTok{{-}1}\NormalTok{) }\OperatorTok{+}
\StringTok{                   }\KeywordTok{nodematch}\NormalTok{(}\StringTok{"race"}\NormalTok{, }\DataTypeTok{diff =} \OtherTok{TRUE}\NormalTok{) }\OperatorTok{+}\StringTok{ }\KeywordTok{absdiff}\NormalTok{(}\StringTok{"sqrt.age.adj"}\NormalTok{) }\OperatorTok{+}
\StringTok{                   }\KeywordTok{offset}\NormalTok{(}\KeywordTok{nodematch}\NormalTok{(}\StringTok{"sex"}\NormalTok{, }\DataTypeTok{diff =} \OtherTok{FALSE}\NormalTok{)) }\OperatorTok{+}\StringTok{ }\KeywordTok{offset}\NormalTok{(concurrent),}
                   \DataTypeTok{target.stats =}\NormalTok{ cohab\_TargetStats,}
                   \DataTypeTok{offset.coef =} \KeywordTok{c}\NormalTok{(}\OperatorTok{{-}}\OtherTok{Inf}\NormalTok{, }\OperatorTok{{-}}\OtherTok{Inf}\NormalTok{),}
                   \DataTypeTok{eval.loglik =} \OtherTok{FALSE}\NormalTok{,}
                   \DataTypeTok{constraints =} \OperatorTok{\textasciitilde{}}\KeywordTok{bd}\NormalTok{(}\DataTypeTok{maxout =} \DecValTok{1}\NormalTok{) }\OperatorTok{+}\StringTok{ }\KeywordTok{blocks}\NormalTok{(}\DataTypeTok{attr =} \OperatorTok{\textasciitilde{}}\NormalTok{sex, }\DataTypeTok{levels2 =} \KeywordTok{diag}\NormalTok{(}\OtherTok{TRUE}\NormalTok{, }\DecValTok{2}\NormalTok{)),}
                   \DataTypeTok{control =} \KeywordTok{snctrl}\NormalTok{(}\DataTypeTok{MCMC.prop =} \OperatorTok{\textasciitilde{}}\KeywordTok{strat}\NormalTok{(}\DataTypeTok{attr =} \OperatorTok{\textasciitilde{}}\NormalTok{race, }\DataTypeTok{empirical =} \OtherTok{TRUE}\NormalTok{) }\OperatorTok{+}\StringTok{ }\NormalTok{sparse,}
                                    \DataTypeTok{init.method =} \StringTok{"MPLE"}\NormalTok{, }\DataTypeTok{init.MPLE.samplesize =} \FloatTok{5e7}\NormalTok{,}
                                    \DataTypeTok{MPLE.constraints.ignore =} \OtherTok{TRUE}\NormalTok{, }\DataTypeTok{MCMLE.effectiveSize =} \OtherTok{NULL}\NormalTok{,}
                                    \DataTypeTok{MCMC.burnin =} \FloatTok{5e4}\NormalTok{, }\DataTypeTok{MCMC.interval =} \FloatTok{5e4}\NormalTok{, }\DataTypeTok{MCMC.samplesize =} \DecValTok{7500}\NormalTok{,}
                                    \DataTypeTok{parallel =}\NormalTok{ ncores, }\DataTypeTok{SAN.nsteps =} \FloatTok{5e7}\NormalTok{,}
                                    \DataTypeTok{SAN.prop=}\OperatorTok{\textasciitilde{}}\KeywordTok{strat}\NormalTok{(}\DataTypeTok{attr =} \OperatorTok{\textasciitilde{}}\NormalTok{race, }\DataTypeTok{pmat =}\NormalTok{ cohab\_MixMat) }\OperatorTok{+}\StringTok{ }\NormalTok{sparse))}
\NormalTok{  el \textless{}{-}}\StringTok{ }\KeywordTok{do.call}\NormalTok{(rbind, }\KeywordTok{lapply}\NormalTok{(fit}\OperatorTok{$}\NormalTok{newnetworks, as.edgelist))}
\NormalTok{  attrs \textless{}{-}}\StringTok{ }\KeywordTok{cbind}\NormalTok{(nw }\OperatorTok{\%v\%}\StringTok{ "race"}\NormalTok{, nw }\OperatorTok{\%v\%}\StringTok{ "age"}\NormalTok{)}
  \KeywordTok{colnames}\NormalTok{(attrs) \textless{}{-}}\StringTok{ }\KeywordTok{c}\NormalTok{(}\StringTok{"race"}\NormalTok{, }\StringTok{"age"}\NormalTok{)}
\NormalTok{  tailattrs \textless{}{-}}\StringTok{ }\NormalTok{attrs[el[,}\DecValTok{1}\NormalTok{],]}
\NormalTok{  headattrs \textless{}{-}}\StringTok{ }\NormalTok{attrs[el[,}\DecValTok{2}\NormalTok{],]}
\NormalTok{  attrnames \textless{}{-}}\StringTok{ "race"}
\NormalTok{  levs \textless{}{-}}\StringTok{ }\KeywordTok{sort}\NormalTok{(}\KeywordTok{unique}\NormalTok{(}\KeywordTok{apply}\NormalTok{(attrs[,attrnames,}\DataTypeTok{drop=}\OtherTok{FALSE}\NormalTok{], }\DecValTok{1}\NormalTok{, paste, }\DataTypeTok{collapse =} \StringTok{"."}\NormalTok{)))}
\NormalTok{  tails \textless{}{-}}\StringTok{ }\KeywordTok{factor}\NormalTok{(}\KeywordTok{apply}\NormalTok{(tailattrs[,attrnames,}\DataTypeTok{drop=}\OtherTok{FALSE}\NormalTok{], }\DecValTok{1}\NormalTok{, paste, }\DataTypeTok{collapse =} \StringTok{"."}\NormalTok{), }\DataTypeTok{levels =}\NormalTok{ levs)}
\NormalTok{  heads \textless{}{-}}\StringTok{ }\KeywordTok{factor}\NormalTok{(}\KeywordTok{apply}\NormalTok{(headattrs[,attrnames,}\DataTypeTok{drop=}\OtherTok{FALSE}\NormalTok{], }\DecValTok{1}\NormalTok{, paste, }\DataTypeTok{collapse =} \StringTok{"."}\NormalTok{), }\DataTypeTok{levels =}\NormalTok{ levs)}
\NormalTok{  mmr \textless{}{-}}\StringTok{ }\KeywordTok{table}\NormalTok{(}\DataTypeTok{from =}\NormalTok{ tails, }\DataTypeTok{to =}\NormalTok{ heads)}
\NormalTok{  mmr \textless{}{-}}\StringTok{ }\NormalTok{mmr }\OperatorTok{+}\StringTok{ }\KeywordTok{t}\NormalTok{(mmr) }\OperatorTok{{-}}\StringTok{ }\KeywordTok{diag}\NormalTok{(}\KeywordTok{diag}\NormalTok{(mmr))}
\NormalTok{  attrnames \textless{}{-}}\StringTok{ }\KeywordTok{c}\NormalTok{(}\StringTok{"race"}\NormalTok{, }\StringTok{"age"}\NormalTok{)}
\NormalTok{  levs \textless{}{-}}\StringTok{ }\KeywordTok{sort}\NormalTok{(}\KeywordTok{unique}\NormalTok{(}\KeywordTok{apply}\NormalTok{(attrs[,attrnames,}\DataTypeTok{drop=}\OtherTok{FALSE}\NormalTok{], }\DecValTok{1}\NormalTok{, paste, }\DataTypeTok{collapse =} \StringTok{"."}\NormalTok{)))}
\NormalTok{  tails \textless{}{-}}\StringTok{ }\KeywordTok{factor}\NormalTok{(}\KeywordTok{apply}\NormalTok{(tailattrs[,attrnames,}\DataTypeTok{drop=}\OtherTok{FALSE}\NormalTok{], }\DecValTok{1}\NormalTok{, paste, }\DataTypeTok{collapse =} \StringTok{"."}\NormalTok{), }\DataTypeTok{levels =}\NormalTok{ levs)}
\NormalTok{  heads \textless{}{-}}\StringTok{ }\KeywordTok{factor}\NormalTok{(}\KeywordTok{apply}\NormalTok{(headattrs[,attrnames,}\DataTypeTok{drop=}\OtherTok{FALSE}\NormalTok{], }\DecValTok{1}\NormalTok{, paste, }\DataTypeTok{collapse =} \StringTok{"."}\NormalTok{), }\DataTypeTok{levels =}\NormalTok{ levs)}
\NormalTok{  mmra \textless{}{-}}\StringTok{ }\KeywordTok{table}\NormalTok{(}\DataTypeTok{from =}\NormalTok{ tails, }\DataTypeTok{to =}\NormalTok{ heads)}
\NormalTok{  mmra \textless{}{-}}\StringTok{ }\NormalTok{mmra }\OperatorTok{+}\StringTok{ }\KeywordTok{t}\NormalTok{(mmra) }\OperatorTok{{-}}\StringTok{ }\KeywordTok{diag}\NormalTok{(}\KeywordTok{diag}\NormalTok{(mmra))}
\NormalTok{  mmra[mmra }\OperatorTok{==}\StringTok{ }\DecValTok{0}\NormalTok{] \textless{}{-}}\StringTok{ }\DecValTok{1}\OperatorTok{/}\DecValTok{2} \CommentTok{\#\# ensure positive proposal weight for all allowed pairings}
\NormalTok{\}}
\end{Highlighting}
\end{Shaded}

The next block of code initializes a Markov chain at an empty
million-node network and tests three different Metropolis--Hastings
proposals, each with equilibrium distribution given by the ERGM
corresponding to the estimated coefficients from above, adjusted as
needed for a network 20 times as large as the network used above. This
code is written to exploit parallel computing and can take a while to
execute, depending on the value of \texttt{nsim}, the required number of
simulated vectors of statistics which, when multiplied by
\texttt{interval}, gives the total number of Metropolis--Hastings
proposals in each Markov chain.

\begin{Shaded}
\begin{Highlighting}[]
\NormalTok{multiplier \textless{}{-}}\StringTok{ }\DecValTok{20}
\ControlFlowTok{if}\NormalTok{ (RunMode }\OperatorTok{!=}\StringTok{ "Skip"}\NormalTok{) \{}
  \KeywordTok{library}\NormalTok{(parallel)}
\NormalTok{  nw \textless{}{-}}\StringTok{ }\KeywordTok{network.initialize}\NormalTok{(net\_size }\OperatorTok{*}\StringTok{ }\NormalTok{multiplier, }\DataTypeTok{directed =} \OtherTok{FALSE}\NormalTok{)}
\NormalTok{  indsLong \textless{}{-}}\StringTok{ }\KeywordTok{rep}\NormalTok{(inds, }\DataTypeTok{length.out =}\NormalTok{ net\_size }\OperatorTok{*}\StringTok{ }\NormalTok{multiplier)}
  \KeywordTok{set.vertex.attribute}\NormalTok{(nw, }\KeywordTok{names}\NormalTok{(cohab\_PopWts)[}\OperatorTok{{-}}\DecValTok{1}\NormalTok{], cohab\_PopWts[indsLong,}\OperatorTok{{-}}\DecValTok{1}\NormalTok{])}
\NormalTok{  coef \textless{}{-}}\StringTok{ }\KeywordTok{coef}\NormalTok{(fit)[}\KeywordTok{seq\_len}\NormalTok{(}\KeywordTok{length}\NormalTok{(}\KeywordTok{coef}\NormalTok{(fit)) }\OperatorTok{{-}}\StringTok{ }\DecValTok{2}\NormalTok{)]}
\NormalTok{  coef[}\DecValTok{1}\NormalTok{] \textless{}{-}}\StringTok{ }\NormalTok{coef[}\DecValTok{1}\NormalTok{] }\OperatorTok{{-}}\StringTok{ }\KeywordTok{log}\NormalTok{(multiplier)}
\NormalTok{  TargetStatsLarge \textless{}{-}}\StringTok{ }\NormalTok{cohab\_TargetStats }\OperatorTok{*}\StringTok{ }\NormalTok{multiplier}
\NormalTok{  attribs \textless{}{-}}\StringTok{ }\KeywordTok{matrix}\NormalTok{(}\OtherTok{FALSE}\NormalTok{, }\DataTypeTok{nrow =} \KeywordTok{network.size}\NormalTok{(nw), }\DataTypeTok{ncol =} \DecValTok{2}\NormalTok{)}
\NormalTok{  attribs[nw }\OperatorTok{\%v\%}\StringTok{ "sex"} \OperatorTok{==}\StringTok{ "M"}\NormalTok{, }\DecValTok{1}\NormalTok{] \textless{}{-}}\StringTok{ }\OtherTok{TRUE}
\NormalTok{  attribs[nw }\OperatorTok{\%v\%}\StringTok{ "sex"} \OperatorTok{==}\StringTok{ "F"}\NormalTok{, }\DecValTok{2}\NormalTok{] \textless{}{-}}\StringTok{ }\OtherTok{TRUE}
\NormalTok{  maxout \textless{}{-}}\StringTok{ }\KeywordTok{matrix}\NormalTok{(}\DecValTok{0}\NormalTok{, }\DataTypeTok{nrow =} \KeywordTok{network.size}\NormalTok{(nw), }\DataTypeTok{ncol =} \DecValTok{2}\NormalTok{)}
\NormalTok{  maxout[nw }\OperatorTok{\%v\%}\StringTok{ "sex"} \OperatorTok{==}\StringTok{ "M"}\NormalTok{, }\DecValTok{2}\NormalTok{] \textless{}{-}}\StringTok{ }\DecValTok{1}
\NormalTok{  maxout[nw }\OperatorTok{\%v\%}\StringTok{ "sex"} \OperatorTok{==}\StringTok{ "F"}\NormalTok{, }\DecValTok{1}\NormalTok{] \textless{}{-}}\StringTok{ }\DecValTok{1}
\NormalTok{  Constraints1 \textless{}{-}}\StringTok{ }\KeywordTok{list}\NormalTok{(}\StringTok{"TNT"}\OperatorTok{\textasciitilde{}}\KeywordTok{bd}\NormalTok{(}\DataTypeTok{attribs =}\NormalTok{ attribs, }\DataTypeTok{maxout =}\NormalTok{ maxout),}
                   \OperatorTok{\textasciitilde{}}\KeywordTok{bd}\NormalTok{(}\DataTypeTok{maxout =} \DecValTok{1}\NormalTok{) }\OperatorTok{+}\StringTok{ }\KeywordTok{blocks}\NormalTok{(}\DataTypeTok{attr =} \StringTok{"sex"}\NormalTok{, }\DataTypeTok{levels2 =} \KeywordTok{diag}\NormalTok{(}\OtherTok{TRUE}\NormalTok{, }\DecValTok{2}\NormalTok{))}
                   \OperatorTok{+}\StringTok{ }\KeywordTok{strat}\NormalTok{(}\DataTypeTok{attr =} \OperatorTok{\textasciitilde{}}\KeywordTok{paste}\NormalTok{(race, }\DataTypeTok{sep =} \StringTok{"."}\NormalTok{), }\DataTypeTok{pmat =}\NormalTok{ mmr),}
                   \OperatorTok{\textasciitilde{}}\KeywordTok{bd}\NormalTok{(}\DataTypeTok{maxout =} \DecValTok{1}\NormalTok{) }\OperatorTok{+}\StringTok{ }\KeywordTok{blocks}\NormalTok{(}\DataTypeTok{attr =} \StringTok{"sex"}\NormalTok{, }\DataTypeTok{levels2 =} \KeywordTok{diag}\NormalTok{(}\OtherTok{TRUE}\NormalTok{, }\DecValTok{2}\NormalTok{))}
                   \OperatorTok{+}\StringTok{ }\KeywordTok{strat}\NormalTok{(}\DataTypeTok{attr =} \OperatorTok{\textasciitilde{}}\KeywordTok{paste}\NormalTok{(race, age, }\DataTypeTok{sep =} \StringTok{"."}\NormalTok{), }\DataTypeTok{pmat =}\NormalTok{ mmra))}
\NormalTok{  ConstraintNames1 \textless{}{-}}\StringTok{ }\KeywordTok{c}\NormalTok{(}\StringTok{"}\CharTok{\textbackslash{}"}\StringTok{TNT}\CharTok{\textbackslash{}"}\StringTok{\textasciitilde{}bd(sex,1)"}\NormalTok{, }\StringTok{"\textasciitilde{}bd(1)+blocks(sex)+strat(race)"}\NormalTok{,}
                 \StringTok{"\textasciitilde{}bd(1)+blocks(sex)+strat(race,age)"}\NormalTok{)}
\NormalTok{  nsim \textless{}{-}}\StringTok{ }\KeywordTok{ifelse}\NormalTok{(RunMode }\OperatorTok{==}\StringTok{"Large"}\NormalTok{, }\FloatTok{1e5}\NormalTok{, }\DecValTok{50}\NormalTok{)}
\NormalTok{  interval \textless{}{-}}\StringTok{ }\FloatTok{1e6}
\NormalTok{  run\_simulate \textless{}{-}}\StringTok{ }\ControlFlowTok{function}\NormalTok{(constraint) \{}
    \KeywordTok{library}\NormalTok{(ergm)}
\NormalTok{    elapsed \textless{}{-}}\StringTok{ }\KeywordTok{system.time}\NormalTok{ (\{}
\NormalTok{      x \textless{}{-}}\StringTok{ }\KeywordTok{simulate}\NormalTok{(CohabFormula, }\DataTypeTok{coef =}\NormalTok{ coef, }\DataTypeTok{constraints =}\NormalTok{ constraint, }\DataTypeTok{output =} \StringTok{"stats"}\NormalTok{,}
                  \DataTypeTok{nsim =}\NormalTok{ nsim, }\DataTypeTok{control =} \KeywordTok{snctrl}\NormalTok{(}\DataTypeTok{MCMC.interval =}\NormalTok{ interval, }\DataTypeTok{MCMC.burnin =}\NormalTok{ interval))}
\NormalTok{    \})}
\NormalTok{    x \textless{}{-}}\StringTok{ }\KeywordTok{matrix}\NormalTok{(}\KeywordTok{c}\NormalTok{(x), }\DataTypeTok{nrow =}\NormalTok{ nsim, }\DataTypeTok{dimnames =} \KeywordTok{dimnames}\NormalTok{(x))}
    \KeywordTok{list}\NormalTok{(}\DataTypeTok{statsmatrix =}\NormalTok{ x, }\DataTypeTok{elapsed=}\NormalTok{elapsed)}
\NormalTok{  \}}
\NormalTok{  cl \textless{}{-}}\StringTok{ }\KeywordTok{makeCluster}\NormalTok{(}\KeywordTok{length}\NormalTok{(Constraints1))}
  \KeywordTok{clusterExport}\NormalTok{(cl, }\StringTok{"nw"}\NormalTok{)}
  \KeywordTok{clusterExport}\NormalTok{(cl, }\StringTok{"CohabFormula"}\NormalTok{)}
  \KeywordTok{clusterExport}\NormalTok{(cl, }\StringTok{"coef"}\NormalTok{)}
  \KeywordTok{clusterExport}\NormalTok{(cl, }\StringTok{"mmr"}\NormalTok{)}
  \KeywordTok{clusterExport}\NormalTok{(cl, }\StringTok{"mmra"}\NormalTok{)}
  \KeywordTok{clusterExport}\NormalTok{(cl, }\StringTok{"nsim"}\NormalTok{)}
  \KeywordTok{clusterExport}\NormalTok{(cl, }\StringTok{"interval"}\NormalTok{)}
  \KeywordTok{clusterExport}\NormalTok{(cl, }\StringTok{"attribs"}\NormalTok{)}
  \KeywordTok{clusterExport}\NormalTok{(cl, }\StringTok{"maxout"}\NormalTok{)}
\NormalTok{  rv \textless{}{-}}\StringTok{ }\KeywordTok{clusterApply}\NormalTok{(cl, Constraints1, run\_simulate)}
  \KeywordTok{stopCluster}\NormalTok{(cl)}
\NormalTok{  z1 \textless{}{-}}\StringTok{ }\KeywordTok{lapply}\NormalTok{(rv, }\StringTok{\textasciigrave{}}\DataTypeTok{[[}\StringTok{\textasciigrave{}}\NormalTok{, }\StringTok{"statsmatrix"}\NormalTok{)}
\NormalTok{  times1 \textless{}{-}}\StringTok{ }\KeywordTok{lapply}\NormalTok{(rv, }\StringTok{\textasciigrave{}}\DataTypeTok{[[}\StringTok{\textasciigrave{}}\NormalTok{, }\StringTok{"elapsed"}\NormalTok{)}
\NormalTok{\}}
\end{Highlighting}
\end{Shaded}

\begin{figure}

{\centering \includegraphics[width=0.95\linewidth]{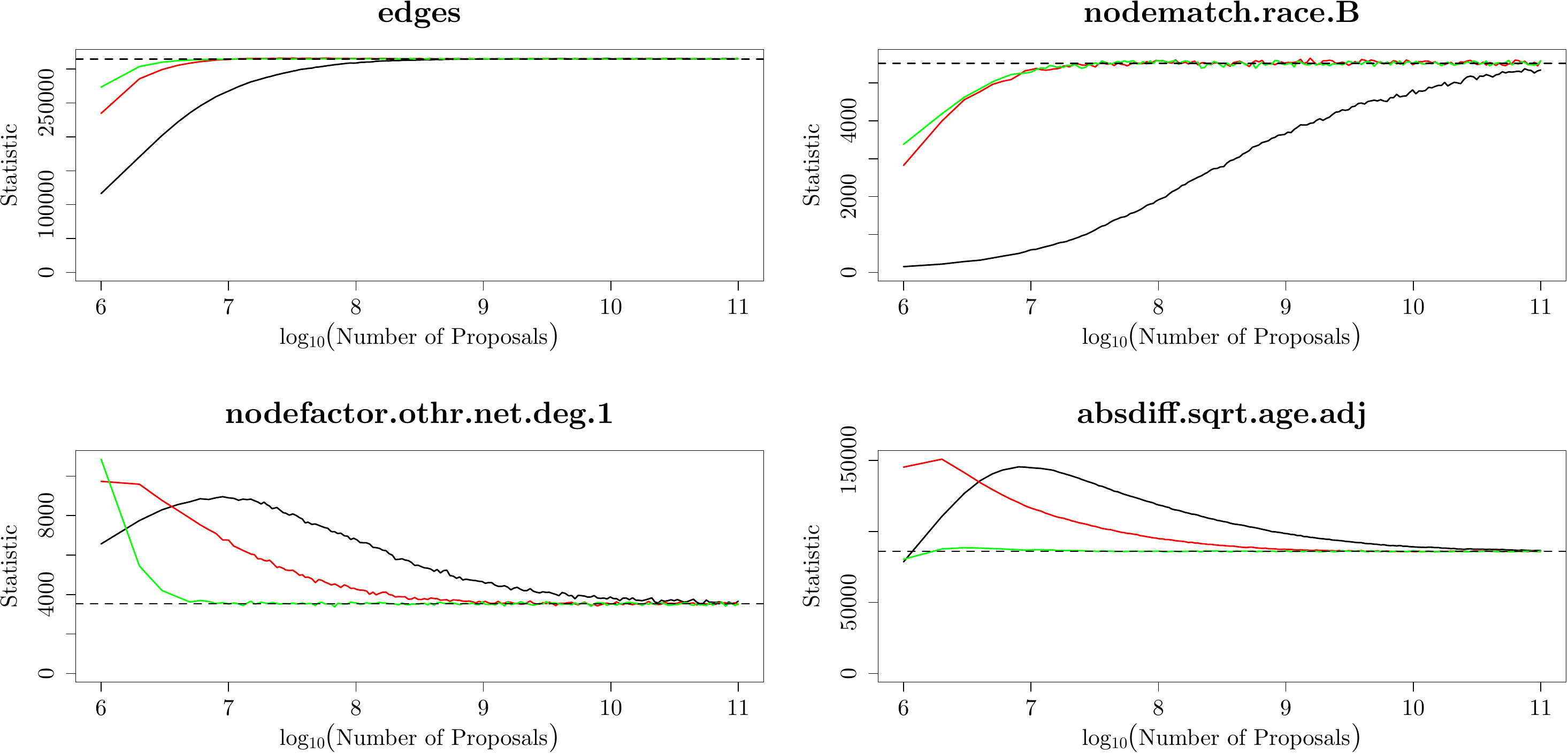} 

}

\caption{Approach to equilibrium of Markov chains using \texttt{TNT} (black), \texttt{BDStratTNT} stratified only on race (red), and \texttt{BDStratTNT} stratified jointly on race and age (green).  Target values are shown as horizontal dotted lines.}\label{fig:MCMCSpeedupFigure}
\end{figure}

The trace plots of \figref{fig:MCMCSpeedupFigure} show how the Markov
chain, starting from the empty network, evolves as a function of the
total number of proposals made. Only four of the fifteen ERGM statistics
are depicted to save space, and the horizontal axis starts at 3 becuase
statistics are sampled only every 1000 proposals. We see that all four
of the statistics depicted have stabilized at their target values after
roughly \(10^7\) proposals using the \texttt{BDStratTNT} proposal
stratified on both race and age. On the other hand, \texttt{TNT} without
any stratification has not yet converged to the target values after
\(10^9\) proposals; in longer tests, we find that even after \(10^{11}\)
proposals the \texttt{nodematch.race.B} statistic has not quite achieved
the target value using \texttt{TNT}.

\figref{fig:MCMCSpeedupFigure} shows that the \texttt{BDStratTNT}
proposal stratifying on only race holds roughly an order of magnitude
advantage over \texttt{TNT} for three of the four statistics shown, and
this advantage increases to 3 or more orders of magnitude for the
statistic representing homophily for race group B
(\texttt{nodematch.race.B}). The \texttt{BDStratTNT} proposal
stratifying on both race and age roughly matches the performance of the
race-only stratification for the statistics representing density
(\texttt{edges}) and homophily for race group B, but it does
significantly better for the terms representing the effects of age
differences (\texttt{absdiff.sqrt.age.adj}) and having at least one
persistent relationship in another network
(\texttt{nodefactor.othr.net.deg.1}).

The faster convergence of the \texttt{absdiff.sqrt.age.adj} statistic
when the proposal considers stratification by age is expected, since the
model favors ties on dyads where the nodes have similar ages and such
dyads will have toggles proposed more frequently when proposals are
stratified according to age mixing.

There is a modest sex asymmetry in the age mixing, with females tending
to be about 1.5 years younger than their male partners; however, taking
this asymmetry into account in the proposal stratification via the
\texttt{pmat} argument of the \texttt{strat} hint does not produce
significant additional gains. The faster convergence of
\texttt{nodefactor.othr.net.deg.1} may arise because nodes having
positive \texttt{othr.net.deg} tend to be younger than the population
average age, so proposal age stratification can hasten equilibration of
the \texttt{nodefactor.othr.net.deg.1} statistic.

\hypertarget{gains-in-effective-sample-size}{%
\subsection{Gains in Effective Sample
Size}\label{gains-in-effective-sample-size}}

\label{sec:ComputingEfficiencyGains}

This section compares Markov chains based on effective sample size
(ESS), as calculated by the \pkg{coda} package (Plummer et al., 2006),
for the 15 individual statistics of the ERGM of
\secref{sec:NewProposals}. The network in these tests has 50,000 nodes,
and the Markov chains are sampled every 100 proposals until a total of
500,000 vectors of statistics are produced. (In longer runs, we have
generated 10 million vectors per chain.) ESS gives a way to compare
various Markov chains that all have the same equilibrium distribution,
since chains that do not mix well do not produce samples that vary much,
which in turn reduces their ESS.

Among the Metropolis--Hastings algorithms we test, some stratify the
nodes by the \texttt{race} variable when proposing node pairs to toggle.
We may use the \texttt{pmat} argument to explicitly pass a matrix of
probabilities that the algorithm should assign to each possible
combination of levels of \texttt{race}, and in our tests we use two
versions of this \texttt{pmat} argument. The first, called \texttt{mmr},
simply assigns probabilities based on the edge fractions observed in the
\texttt{cohab} dataset. The second, called \texttt{mmr\_mod}, gives more
weight, in selecting potential MCMC edge toggles, to smaller race groups
than their edge fraction would dictate: Finally, the \texttt{mmra}
matrix, used for the proposal that stratifies by both \texttt{race} and
\texttt{age}, simply uses the observed edge fractions for each stratum.

\begin{Shaded}
\begin{Highlighting}[]
\ControlFlowTok{if}\NormalTok{ (RunMode }\OperatorTok{!=}\StringTok{ "Skip"}\NormalTok{) \{}
\NormalTok{  mmr\_mod \textless{}{-}}\StringTok{ }\NormalTok{mmr}
\NormalTok{  mmr\_mod[}\OperatorTok{{-}}\NormalTok{(}\DecValTok{4}\OperatorTok{:}\DecValTok{5}\NormalTok{),] \textless{}{-}}\StringTok{ }\NormalTok{mmr\_mod[,}\OperatorTok{{-}}\NormalTok{(}\DecValTok{4}\OperatorTok{:}\DecValTok{5}\NormalTok{)]}\OperatorTok{*}\KeywordTok{sqrt}\NormalTok{(}\DecValTok{6}\NormalTok{)}
\NormalTok{  mmr\_mod[,}\OperatorTok{{-}}\NormalTok{(}\DecValTok{4}\OperatorTok{:}\DecValTok{5}\NormalTok{)] \textless{}{-}}\StringTok{ }\NormalTok{mmr\_mod[}\OperatorTok{{-}}\NormalTok{(}\DecValTok{4}\OperatorTok{:}\DecValTok{5}\NormalTok{),]}\OperatorTok{*}\KeywordTok{sqrt}\NormalTok{(}\DecValTok{6}\NormalTok{)}
\NormalTok{  mmr\_mod[}\DecValTok{3}\NormalTok{,] \textless{}{-}}\StringTok{ }\NormalTok{mmr\_mod[}\DecValTok{3}\NormalTok{,]}\OperatorTok{*}\KeywordTok{sqrt}\NormalTok{(}\DecValTok{2}\NormalTok{)}
\NormalTok{  mmr\_mod[,}\DecValTok{3}\NormalTok{] \textless{}{-}}\StringTok{ }\NormalTok{mmr\_mod[,}\DecValTok{3}\NormalTok{]}\OperatorTok{*}\KeywordTok{sqrt}\NormalTok{(}\DecValTok{2}\NormalTok{)}
\NormalTok{  mmr\_mod[}\DecValTok{4}\NormalTok{,] \textless{}{-}}\StringTok{ }\FloatTok{1.5}\OperatorTok{*}\NormalTok{mmr\_mod[}\DecValTok{4}\NormalTok{,]}
\NormalTok{  mmr\_mod[,}\DecValTok{4}\NormalTok{] \textless{}{-}}\StringTok{ }\FloatTok{1.5}\OperatorTok{*}\NormalTok{mmr\_mod[,}\DecValTok{4}\NormalTok{]}
\NormalTok{\}}
\end{Highlighting}
\end{Shaded}

Now that the network and all necessary supporting objects are in place,
we can run the test. The value of \texttt{nsimESS} gives the number of
15-dimensional vectors of statistics, sampled once every 100 proposals
according to the value of \texttt{interval}, produced before the Markov
chain is stopped.

\begin{Shaded}
\begin{Highlighting}[]
\NormalTok{nsimESS \textless{}{-}}\StringTok{ }\KeywordTok{ifelse}\NormalTok{(RunMode }\OperatorTok{==}\StringTok{"Small"}\NormalTok{, }\FloatTok{1e5}\NormalTok{, }\FloatTok{1e7}\NormalTok{)}
\NormalTok{intervalESS \textless{}{-}}\StringTok{ }\DecValTok{100}
\ControlFlowTok{if}\NormalTok{ (RunMode }\OperatorTok{!=}\StringTok{ "Skip"}\NormalTok{) \{}
\NormalTok{  nw \textless{}{-}}\StringTok{ }\NormalTok{fit}\OperatorTok{$}\NormalTok{newnetworks[[}\DecValTok{1}\NormalTok{]] }\CommentTok{\# Use network from previous 50000{-}node fit}
\NormalTok{  coef \textless{}{-}}\StringTok{ }\KeywordTok{coef}\NormalTok{(fit)[}\KeywordTok{seq\_len}\NormalTok{(}\KeywordTok{length}\NormalTok{(}\KeywordTok{coef}\NormalTok{(fit)) }\OperatorTok{{-}}\StringTok{ }\DecValTok{2}\NormalTok{)] }\CommentTok{\# Use estimates from previous fit}
\CommentTok{\#  save(nw, coef, file="nwANDcoef.RData")}
\CommentTok{\#\} }
\NormalTok{  attribs \textless{}{-}}\StringTok{ }\KeywordTok{matrix}\NormalTok{(}\OtherTok{FALSE}\NormalTok{, }\DataTypeTok{nrow =} \KeywordTok{network.size}\NormalTok{(nw), }\DataTypeTok{ncol =} \DecValTok{2}\NormalTok{)}
\NormalTok{  attribs[nw }\OperatorTok{\%v\%}\StringTok{ "sex"} \OperatorTok{==}\StringTok{ "M"}\NormalTok{, }\DecValTok{1}\NormalTok{] \textless{}{-}}\StringTok{ }\OtherTok{TRUE}
\NormalTok{  attribs[nw }\OperatorTok{\%v\%}\StringTok{ "sex"} \OperatorTok{==}\StringTok{ "F"}\NormalTok{, }\DecValTok{2}\NormalTok{] \textless{}{-}}\StringTok{ }\OtherTok{TRUE}
\NormalTok{  maxout \textless{}{-}}\StringTok{ }\KeywordTok{matrix}\NormalTok{(}\DecValTok{0}\NormalTok{, }\DataTypeTok{nrow =} \KeywordTok{network.size}\NormalTok{(nw), }\DataTypeTok{ncol =} \DecValTok{2}\NormalTok{)}
\NormalTok{  maxout[nw }\OperatorTok{\%v\%}\StringTok{ "sex"} \OperatorTok{==}\StringTok{ "M"}\NormalTok{, }\DecValTok{2}\NormalTok{] \textless{}{-}}\StringTok{ }\DecValTok{1}
\NormalTok{  maxout[nw }\OperatorTok{\%v\%}\StringTok{ "sex"} \OperatorTok{==}\StringTok{ "F"}\NormalTok{, }\DecValTok{1}\NormalTok{] \textless{}{-}}\StringTok{ }\DecValTok{1}
\NormalTok{  Constraints2 \textless{}{-}}\StringTok{ }\KeywordTok{list}\NormalTok{(}\StringTok{"TNT"}\OperatorTok{\textasciitilde{}}\KeywordTok{bd}\NormalTok{(}\DataTypeTok{attribs =}\NormalTok{ attribs, }\DataTypeTok{maxout =}\NormalTok{ maxout),}
                 \OperatorTok{\textasciitilde{}}\KeywordTok{bd}\NormalTok{(}\DataTypeTok{maxout =} \DecValTok{1}\NormalTok{) }\OperatorTok{+}\StringTok{ }\KeywordTok{blocks}\NormalTok{(}\DataTypeTok{attr =} \StringTok{"sex"}\NormalTok{, }\DataTypeTok{levels2 =} \KeywordTok{diag}\NormalTok{(}\OtherTok{TRUE}\NormalTok{, }\DecValTok{2}\NormalTok{)),}
                 \OperatorTok{\textasciitilde{}}\KeywordTok{bd}\NormalTok{(}\DataTypeTok{maxout =} \DecValTok{1}\NormalTok{) }\OperatorTok{+}\StringTok{ }\KeywordTok{blocks}\NormalTok{(}\DataTypeTok{attr =} \StringTok{"sex"}\NormalTok{, }\DataTypeTok{levels2 =} \KeywordTok{diag}\NormalTok{(}\OtherTok{TRUE}\NormalTok{, }\DecValTok{2}\NormalTok{))}
                                 \OperatorTok{+}\StringTok{ }\KeywordTok{strat}\NormalTok{(}\DataTypeTok{attr =} \OperatorTok{\textasciitilde{}}\KeywordTok{paste}\NormalTok{(race, }\DataTypeTok{sep =} \StringTok{"."}\NormalTok{), }\DataTypeTok{pmat =}\NormalTok{ mmr),}
                 \OperatorTok{\textasciitilde{}}\KeywordTok{bd}\NormalTok{(}\DataTypeTok{maxout =} \DecValTok{1}\NormalTok{) }\OperatorTok{+}\StringTok{ }\KeywordTok{blocks}\NormalTok{(}\DataTypeTok{attr =} \StringTok{"sex"}\NormalTok{, }\DataTypeTok{levels2 =} \KeywordTok{diag}\NormalTok{(}\OtherTok{TRUE}\NormalTok{, }\DecValTok{2}\NormalTok{))}
                                 \OperatorTok{+}\StringTok{ }\KeywordTok{strat}\NormalTok{(}\DataTypeTok{attr =} \OperatorTok{\textasciitilde{}}\KeywordTok{paste}\NormalTok{(race, }\DataTypeTok{sep =} \StringTok{"."}\NormalTok{), }\DataTypeTok{pmat =}\NormalTok{ mmr\_mod),}
                 \OperatorTok{\textasciitilde{}}\KeywordTok{bd}\NormalTok{(}\DataTypeTok{maxout =} \DecValTok{1}\NormalTok{) }\OperatorTok{+}\StringTok{ }\KeywordTok{blocks}\NormalTok{(}\DataTypeTok{attr =} \StringTok{"sex"}\NormalTok{, }\DataTypeTok{levels2 =} \KeywordTok{diag}\NormalTok{(}\OtherTok{TRUE}\NormalTok{, }\DecValTok{2}\NormalTok{))}
                                 \OperatorTok{+}\StringTok{ }\KeywordTok{strat}\NormalTok{(}\DataTypeTok{attr =} \OperatorTok{\textasciitilde{}}\KeywordTok{paste}\NormalTok{(race, age, }\DataTypeTok{sep =} \StringTok{"."}\NormalTok{), }\DataTypeTok{pmat =}\NormalTok{ mmra))}
\NormalTok{  ConstraintNames2 \textless{}{-}}\StringTok{ }\KeywordTok{c}\NormalTok{(}\StringTok{"}\CharTok{\textbackslash{}"}\StringTok{TNT}\CharTok{\textbackslash{}"}\StringTok{\textasciitilde{}bd(sex,1)"}\NormalTok{,}
               \StringTok{"\textasciitilde{}bd(1)+blocks(sex)"}\NormalTok{,}
               \StringTok{"\textasciitilde{}bd(1)+blocks(sex)+strat(race)"}\NormalTok{,}
               \StringTok{"\textasciitilde{}bd(1)+blocks(sex)+strat(race.mod)"}\NormalTok{,}
               \StringTok{"\textasciitilde{}bd(1)+blocks(sex)+strat(race,age)"}\NormalTok{)}
  \KeywordTok{set.seed}\NormalTok{(}\DecValTok{0}\NormalTok{)}
\NormalTok{  z2 \textless{}{-}}\StringTok{ }\KeywordTok{matrix}\NormalTok{(}\DecValTok{0}\NormalTok{, }\DecValTok{5}\NormalTok{, }\DecValTok{15}\NormalTok{)}
  \KeywordTok{rownames}\NormalTok{(z2) \textless{}{-}}\StringTok{ }\NormalTok{ConstraintNames2}
\NormalTok{  times2 \textless{}{-}}\StringTok{ }\KeywordTok{list}\NormalTok{()}
  \ControlFlowTok{for}\NormalTok{(i }\ControlFlowTok{in} \KeywordTok{seq\_along}\NormalTok{(Constraints2)) \{}
\NormalTok{    times2[[i]] \textless{}{-}}\StringTok{ }\KeywordTok{system.time}\NormalTok{(\{}
\NormalTok{        x \textless{}{-}}\StringTok{ }\KeywordTok{simulate}\NormalTok{(CohabFormula,}
                  \DataTypeTok{coef =}\NormalTok{ coef,}
                  \DataTypeTok{constraints =}\NormalTok{ Constraints2[[i]],}
                  \DataTypeTok{nsim =}\NormalTok{ nsimESS,}
                  \DataTypeTok{output =} \StringTok{"stats"}\NormalTok{,}
                  \DataTypeTok{control =} \KeywordTok{snctrl}\NormalTok{(}\DataTypeTok{MCMC.interval =}\NormalTok{ intervalESS,}
                                   \DataTypeTok{MCMC.burnin =}\NormalTok{ intervalESS))}
\NormalTok{    z2[i,] \textless{}{-}}\StringTok{ }\NormalTok{coda}\OperatorTok{::}\KeywordTok{effectiveSize}\NormalTok{(x)}
\NormalTok{    \})}
\NormalTok{  \}}
\NormalTok{\}}
\end{Highlighting}
\end{Shaded}

The first column in \tabref{tab:ESStable} indicates, in abbreviated
form, the hints and/or constraints that were specified, in addition to
the default \texttt{sparse} hint used for all rows. Thus, the first row
uses the \texttt{TNT} proposal, while all other rows use
\texttt{BDStratTNT} with varying levels of complexity in the hints and
constraints passed to the proposal.

\begin{table}

\caption{\label{tab:ESStable}Effective sample sizes for 4 of the 15 statistics from an MCMC sample of size 1e+07 with interval 100 using the hints and constraints shown in the leftmost column in addition to the \texttt{sparse} hint, which is used in all cases.}
\centering
\begin{tabular}[t]{l>{\raggedleft\arraybackslash}p{4em}>{\raggedleft\arraybackslash}p{4em}>{\raggedleft\arraybackslash}p{4em}>{\raggedleft\arraybackslash}p{4em}}
\toprule
  & edges & race B homoph. & other net deg. 1+ & $\sqrt{\text{age diff.}}$\\
\midrule
\code{\code{"TNT"~bd(sex,1)}} & 2862.0 & 207.8 & 5495.6 & 2149.4\\
\code{\code{~bd(1)+blocks(sex)}} & 25720.3 & 1016.7 & 10671.4 & 12228.1\\
\code{\code{~bd(1)+blocks(sex)+strat(race)}} & 27421.0 & 4336.6 & 11706.8 & 13732.8\\
\code{\code{~bd(1)+blocks(sex)+strat(race.mod)}} & 20206.4 & 12715.2 & 10565.9 & 10360.8\\
\code{\code{~bd(1)+blocks(sex)+strat(race,age)}} & 28086.4 & 4941.7 & 15569.6 & 6892.2\\
\bottomrule
\end{tabular}
\end{table}

Of the proposals tested, only \texttt{TNT} was available in \pkg{ergm}
prior to version 3.10. The various versions of the \texttt{BDStratTNT}
proposal all produce larger ESS values for every statistic measured. If
we compare the first row of \tabref{tab:ESStable} with, say, the fourth
row, representing the \texttt{BDStratTNT} proposal that stratifies on
the modified \texttt{race} effect and respects the heterosexual nature
of the model and the bound placed on degree---no node is ever allowed to
have more than one tie---we see that the smallest value in row 4, 570.4,
is roughly 70 times as large as the smallest value in row 1, 8.2.
Indeed, these two ESS values are the smallest in their respective rows
across all fifteen statistics:

\begin{Shaded}
\begin{Highlighting}[]
\KeywordTok{min}\NormalTok{(z2[}\DecValTok{4}\NormalTok{, ])}\OperatorTok{/}\KeywordTok{min}\NormalTok{(z2[}\DecValTok{1}\NormalTok{, ])}
\end{Highlighting}
\end{Shaded}

\begin{verbatim}
## [1] 95.2256
\end{verbatim}

Since different proposals require different computing effort, we may
also compare the proposals by dividing ESS by the total time required.
\tabref{tab:ESSperMinute} summarizes these ESS per minute measurements.
Comparing rows 1 and 4 as before, we see that the improvement in minimum
ESS per minute across all 15 statistics is roughly 86-fold.

\begin{table}

\caption{\label{tab:ESSperMinute}Effective sample sizes per minute for 4 of the 15 statistics from an MCMC sample of size 1e+07 with interval 100 using the hints and constraints shown in the leftmost column in addition to the \texttt{sparse} hint, which is used in all cases.}
\centering
\begin{tabular}[t]{l>{\raggedleft\arraybackslash}p{4em}>{\raggedleft\arraybackslash}p{4em}>{\raggedleft\arraybackslash}p{4em}>{\raggedleft\arraybackslash}p{4em}}
\toprule
  & edges & race B homoph. & other net deg. 1+ & $\sqrt{\text{age diff.}}$\\
\midrule
\code{"TNT"~bd(sex,1)} & 310.2 & 22.5 & 595.7 & 233.0\\
\code{~bd(1)+blocks(sex)} & 3397.2 & 134.3 & 1409.5 & 1615.1\\
\code{~bd(1)+blocks(sex)+strat(race)} & 3401.1 & 537.9 & 1452.0 & 1703.3\\
\code{~bd(1)+blocks(sex)+strat(race.mod)} & 2461.8 & 1549.2 & 1287.3 & 1262.3\\
\code{~bd(1)+blocks(sex)+strat(race,age)} & 2185.4 & 384.5 & 1211.5 & 536.3\\
\bottomrule
\end{tabular}
\end{table}

\hypertarget{impact-of-mcmc-improvements-on-simulated-annealing-speed}{%
\subsection{Impact of MCMC improvements on simulated annealing
speed}\label{impact-of-mcmc-improvements-on-simulated-annealing-speed}}

\label{sec:SANspeedup}

\begin{figure}

{\centering \includegraphics[width=0.95\linewidth]{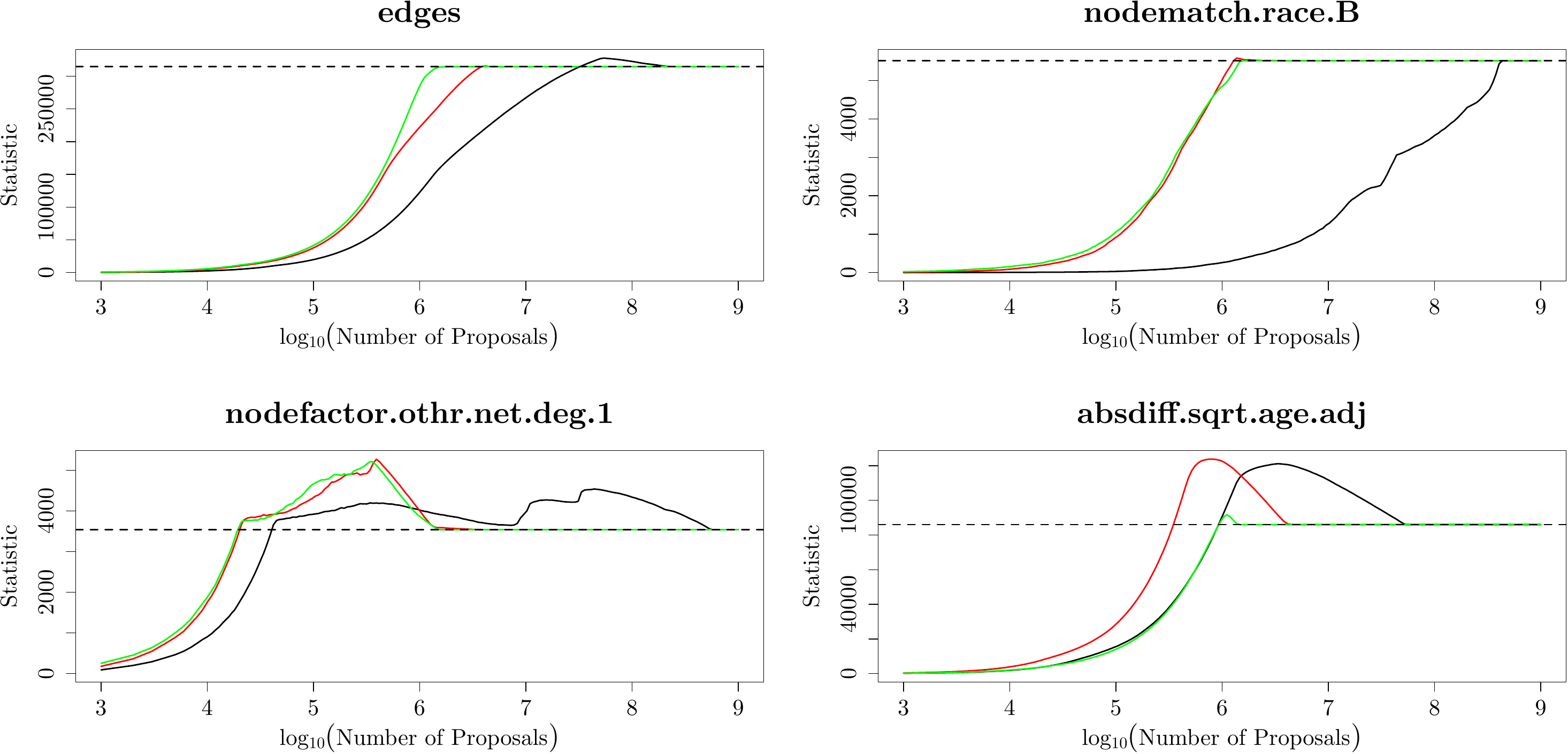} 

}

\caption{Approach to target values of SAN runs using proposals \texttt{TNT} (black), \texttt{BDStratTNT} stratified only on race (red), and \texttt{BDStratTNT} stratified jointly on race and age (green).  Target values are shown as horizontal lines.}\label{fig:SANStratification}
\end{figure}

The trace plots in \figref{fig:SANStratification} show how various
statistics approach their target values during a run of \pkg{ergm}'s
simulated annealing algorithm, starting from an empty million-node
network, with each of three different proposals. The horizontal axis is
the base 10 logarithm of the number of proposals made, and the vertical
axis is the statistic value, with the target value indicated as the
horizontal purple line. Statistics are sampled every 1000 proposals, so
the horizontal axis starts at 3. The SAN run takes place at a fixed
temperature of 0, with the matrix of weights being the diagonal matrix
of reciprocal squared target statistics divided by their sum. This
choice of temperature and weight matrix settings was made to try to
minimize the effect of these choices on the algorithm's behavior and
thereby isolate the different effects of the proposals, yet we find that
using the default TNT settings explained in \secref{sec:SANalgorithm}
leads to similar behavior to that reported in
\figref{fig:SANStratification}.

\figref{fig:SANStratification} shows that the \texttt{BDStratTNT}
proposal stratifying on only race yields an advantage of roughly 1 to 2
orders of magnitude over \texttt{TNT} for these statistics, with the
\texttt{BDStratTNT} proposal stratifying on both race and age yielding
an additional half an order of magnitude or less. Here is the code used
to produce the figure:

\begin{Shaded}
\begin{Highlighting}[]
\ControlFlowTok{if}\NormalTok{ (RunMode }\OperatorTok{!=}\StringTok{ "Skip"}\NormalTok{) \{}
  \KeywordTok{library}\NormalTok{(parallel)}
  \KeywordTok{set.seed}\NormalTok{(}\DecValTok{0}\NormalTok{)}
\NormalTok{  nw \textless{}{-}}\StringTok{ }\KeywordTok{network.initialize}\NormalTok{(net\_size }\OperatorTok{*}\StringTok{ }\NormalTok{multiplier, }\DataTypeTok{directed =} \OtherTok{FALSE}\NormalTok{)}
  \KeywordTok{set.vertex.attribute}\NormalTok{(nw, }\KeywordTok{names}\NormalTok{(cohab\_PopWts)[}\OperatorTok{{-}}\DecValTok{1}\NormalTok{], cohab\_PopWts[indsLong,}\OperatorTok{{-}}\DecValTok{1}\NormalTok{])}
\NormalTok{  attribs \textless{}{-}}\StringTok{ }\KeywordTok{matrix}\NormalTok{(}\OtherTok{FALSE}\NormalTok{, }\DataTypeTok{nrow =} \KeywordTok{network.size}\NormalTok{(nw), }\DataTypeTok{ncol =} \DecValTok{2}\NormalTok{)}
\NormalTok{  attribs[nw }\OperatorTok{\%v\%}\StringTok{ "sex"} \OperatorTok{==}\StringTok{ "M"}\NormalTok{, }\DecValTok{1}\NormalTok{] \textless{}{-}}\StringTok{ }\OtherTok{TRUE}
\NormalTok{  attribs[nw }\OperatorTok{\%v\%}\StringTok{ "sex"} \OperatorTok{==}\StringTok{ "F"}\NormalTok{, }\DecValTok{2}\NormalTok{] \textless{}{-}}\StringTok{ }\OtherTok{TRUE}
\NormalTok{  maxout \textless{}{-}}\StringTok{ }\KeywordTok{matrix}\NormalTok{(}\DecValTok{0}\NormalTok{, }\DataTypeTok{nrow =} \KeywordTok{network.size}\NormalTok{(nw), }\DataTypeTok{ncol =} \DecValTok{2}\NormalTok{)}
\NormalTok{  maxout[nw }\OperatorTok{\%v\%}\StringTok{ "sex"} \OperatorTok{==}\StringTok{ "M"}\NormalTok{, }\DecValTok{2}\NormalTok{] \textless{}{-}}\StringTok{ }\DecValTok{1}
\NormalTok{  maxout[nw }\OperatorTok{\%v\%}\StringTok{ "sex"} \OperatorTok{==}\StringTok{ "F"}\NormalTok{, }\DecValTok{1}\NormalTok{] \textless{}{-}}\StringTok{ }\DecValTok{1}
\NormalTok{  samplesize \textless{}{-}}\StringTok{ }\KeywordTok{ifelse}\NormalTok{(RunMode  }\OperatorTok{==}\StringTok{"Large"}\NormalTok{, }\FloatTok{1e6}\NormalTok{, }\FloatTok{5e4}\NormalTok{)}
\NormalTok{  nsteps \textless{}{-}}\StringTok{ }\KeywordTok{ifelse}\NormalTok{(RunMode  }\OperatorTok{==}\StringTok{"Large"}\NormalTok{, }\FloatTok{1e9}\NormalTok{, }\FloatTok{5e7}\NormalTok{)}
\NormalTok{  invcov \textless{}{-}}\StringTok{ }\KeywordTok{diag}\NormalTok{(}\DecValTok{1}\OperatorTok{/}\NormalTok{(TargetStatsLarge}\OperatorTok{**}\DecValTok{2}\NormalTok{))}
\NormalTok{  invcov \textless{}{-}}\StringTok{ }\NormalTok{invcov}\OperatorTok{/}\KeywordTok{sum}\NormalTok{(invcov)}
\NormalTok{  run\_san \textless{}{-}}\StringTok{ }\ControlFlowTok{function}\NormalTok{(constraint) \{}
    \KeywordTok{library}\NormalTok{(ergm)}
\NormalTok{    elapsed \textless{}{-}}\StringTok{ }\KeywordTok{system.time}\NormalTok{(\{}
\NormalTok{      rv \textless{}{-}}\StringTok{ }\KeywordTok{san}\NormalTok{(CohabFormula,}
              \DataTypeTok{constraints =}\NormalTok{ constraint,}
              \DataTypeTok{target.stats =}\NormalTok{ TargetStatsLarge,}
              \DataTypeTok{control =} \KeywordTok{snctrl}\NormalTok{(}\DataTypeTok{SAN.maxit =} \DecValTok{1}\NormalTok{,}
                               \DataTypeTok{SAN.invcov =}\NormalTok{ invcov,}
                               \DataTypeTok{SAN.nsteps =}\NormalTok{ nsteps,}
                               \DataTypeTok{SAN.samplesize =}\NormalTok{ samplesize))}
\NormalTok{    \})}
\NormalTok{    sm \textless{}{-}}\StringTok{ }\KeywordTok{attr}\NormalTok{(rv, }\StringTok{"stats"}\NormalTok{)}
\NormalTok{    sm \textless{}{-}}\StringTok{ }\KeywordTok{t}\NormalTok{(}\KeywordTok{t}\NormalTok{(sm) }\OperatorTok{+}\StringTok{ }\NormalTok{TargetStatsLarge)}
    \KeywordTok{list}\NormalTok{(}\DataTypeTok{statsmatrix =}\NormalTok{ sm, }\DataTypeTok{elapsed =}\NormalTok{ elapsed)}
\NormalTok{  \}}
\NormalTok{  cl \textless{}{-}}\StringTok{ }\KeywordTok{makeCluster}\NormalTok{(}\KeywordTok{length}\NormalTok{(Constraints1))}
  \KeywordTok{clusterExport}\NormalTok{(cl, }\StringTok{"CohabFormula"}\NormalTok{)}
  \KeywordTok{clusterExport}\NormalTok{(cl, }\StringTok{"nw"}\NormalTok{)}
  \KeywordTok{clusterExport}\NormalTok{(cl, }\StringTok{"mmr"}\NormalTok{)}
  \KeywordTok{clusterExport}\NormalTok{(cl, }\StringTok{"mmra"}\NormalTok{)}
  \KeywordTok{clusterExport}\NormalTok{(cl, }\StringTok{"samplesize"}\NormalTok{)}
  \KeywordTok{clusterExport}\NormalTok{(cl, }\StringTok{"nsteps"}\NormalTok{)}
  \KeywordTok{clusterExport}\NormalTok{(cl, }\StringTok{"invcov"}\NormalTok{)}
  \KeywordTok{clusterExport}\NormalTok{(cl, }\StringTok{"TargetStatsLarge"}\NormalTok{)}
  \KeywordTok{clusterExport}\NormalTok{(cl, }\StringTok{"attribs"}\NormalTok{)}
  \KeywordTok{clusterExport}\NormalTok{(cl, }\StringTok{"maxout"}\NormalTok{)}
\NormalTok{  rv \textless{}{-}}\StringTok{ }\KeywordTok{clusterApply}\NormalTok{(cl, Constraints1, run\_san)}
  \KeywordTok{stopCluster}\NormalTok{(cl)}
\NormalTok{  z3 \textless{}{-}}\StringTok{ }\KeywordTok{lapply}\NormalTok{(rv, }\StringTok{\textasciigrave{}}\DataTypeTok{[[}\StringTok{\textasciigrave{}}\NormalTok{, }\StringTok{"statsmatrix"}\NormalTok{)}
\NormalTok{  times3 \textless{}{-}}\StringTok{ }\KeywordTok{lapply}\NormalTok{(rv, }\StringTok{\textasciigrave{}}\DataTypeTok{[[}\StringTok{\textasciigrave{}}\NormalTok{, }\StringTok{"elapsed"}\NormalTok{)}
\NormalTok{\}}
\end{Highlighting}
\end{Shaded}

\hypertarget{impact-of-mcmc-improvements-on-estimation-time}{%
\subsection{Impact of MCMC improvements on estimation
time}\label{impact-of-mcmc-improvements-on-estimation-time}}

\label{sec:EstimationTime}

The cumulative effect of the myriad improvements to the computing
machinery of the \pkg{ergm} package are perhaps best appreciated by
comparing version 4 with earlier versions of the package.
\tabref{tab:Estimationtable} shows computing times for fitting the model
described in \secref{sec:NewProposals} to a network with 50,000 nodes.
The two versions of \pkg{ergm} are the latest version 4 as well as
version 3.10, which may be obtained from CRAN archives. While version
3.10 was released in 2019 and is therefore quite a bit more recent than
the version (2.1) that was published along with Hunter et al. (2008),
the efficiency improvements are nonetheless substantial for this
particular model-fitting example.

For the MCMC proposals, the version 4 runs use \texttt{BDStratTNT},
while the proposals that allow for stratification while taking bounded
degree constraints into account were not available for version 3.10;
thus, the 3.10 runs use plain \texttt{TNT}. All the runs enforce network
constraints via offset terms with coefficients set to \(-\infty\) when
maximizing the pseudolikelihood, as the MPLE procedure cannot handle
such constraints. The ``3.10 without \texttt{bd}'' fits also utilize
these offsets to enforce constraints during MCMC, while the offsets are
redundant in MCMC for the ``3.10 with \texttt{bd}'' and version 4 fits.
An important difference between ``3.10 with \texttt{bd}'' and version 4,
which also uses \texttt{bd}, is the implementation of \texttt{bd}: in
\pkg{ergm} 3.10, the only \texttt{bd} implementation available was a
rejection algorithm, while in \pkg{ergm} 4, the \texttt{BDStratTNT}
proposal maintains the necessary state to avoid the need for a rejection
algorithm when imposing upper bounds on degree.

In this particular model, we have a target mean degree of about 0.63,
meaning that approximately 86\% of randomly chosen dyads cannot be
toggled without violating the degree bound when the network is near
equilibrium. Of the 14\% that can, half cannot be toggled due to the
heterosexuality constraint, which is also taken into account by
\texttt{BDStratTNT}. The \texttt{BDStratTNT} proposal is thus naively
about 15 times as efficient for this model as a proposal that does not
constructively take the constraints into account. The stratification of
proposals by \texttt{race}, also handled by \texttt{BDStratTNT}, yields
still further improvement. This expected baseline increase in
efficiency, together with the effective sample size results described in
\ref{sec:ComputingEfficiencyGains}, explains the choice of fixed
interval values 1/20 as large in the third row of
\tabref{tab:Estimationtable} as in the first two rows. The rightmost
column of \tabref{tab:Estimationtable} uses adaptive MCMC to fit the
model, which is new in \pkg{ergm} 4 and is the default. One may disable
adaptive MCMC by setting \texttt{MCMLE.effectiveSize\ =\ NULL}.

All fits reported in the table parallelize MCMC using an iMac with 6
cores, so improvements in parallelization between \pkg{ergm} versions
3.10 and 4 are also represented in \tabref{tab:Estimationtable}.

A few words are in order regarding the \texttt{bd} constraint and its
relationship with simulated annealing in earlier versions of \pkg{ergm}.
As mentioned above, it is possible to enforce a constraint such as a
bound on degree without using \texttt{bd} by adding an offset term to
the model, say, \texttt{degree(k)} where \texttt{k} is one larger than
the maximum allowable degree, and fixing its coefficient value at
\texttt{-Inf}. However, offsets were previously ignored by \texttt{san},
the simulated annealing function used to produce an initial network,
potentially resulting in an initial network not satisfying the
constraints. This in turn could produce a poor initial parameter value,
as obtained from the MPLE of the network generated by \texttt{san}.
Indeed, we see in the first row of \tabref{tab:Estimationtable} that
models fit in \texttt{3.10} without using \texttt{bd} produce much
longer fit times for this particular model. The \texttt{san} function
now respects offsets, but the facts that multiple algorithms in
\pkg{ergm} might rely on constraints and not all such algorithms
currently optimize their treatment of both offsets and explicit
\texttt{constraints} lead us to recommend the redundancy in specifying
such constraints.

First, we present the code used to test version 4. There is no
\texttt{"Small"} option for the tests in this subsection; using
\texttt{RunMode=="Small"} employs the results from a previously saved
run, whereas \texttt{RunMode=="Large"} takes a very long time, possibly
several days, to finish.

\begin{Shaded}
\begin{Highlighting}[]
\ControlFlowTok{if}\NormalTok{ (RunMode }\OperatorTok{==}\StringTok{ "Large"}\NormalTok{) \{ }\CommentTok{\# There is no "Small" option here; "Small" works like "Skip"}
\NormalTok{  net\_size \textless{}{-}}\StringTok{ }\FloatTok{5e4}
\NormalTok{  V4Times \textless{}{-}}\StringTok{ }\KeywordTok{list}\NormalTok{()}
\NormalTok{  reps \textless{}{-}}\StringTok{ }\KeywordTok{ifelse}\NormalTok{(RunMode }\OperatorTok{==}\StringTok{"Large"}\NormalTok{, }\DecValTok{5}\NormalTok{, }\DecValTok{1}\NormalTok{)}
  \ControlFlowTok{for}\NormalTok{(i }\ControlFlowTok{in} \KeywordTok{seq\_len}\NormalTok{(reps)) \{ }\CommentTok{\# number of repetitions}
    \KeywordTok{set.seed}\NormalTok{(i)}
    \ControlFlowTok{for}\NormalTok{ (j }\ControlFlowTok{in} \DecValTok{1}\OperatorTok{:}\DecValTok{3}\NormalTok{) \{ }\CommentTok{\# first two nonadaptive; third adaptive}
\NormalTok{      nw \textless{}{-}}\StringTok{ }\KeywordTok{network.initialize}\NormalTok{(net\_size, }\DataTypeTok{directed =} \OtherTok{FALSE}\NormalTok{)}
\NormalTok{      inds \textless{}{-}}\StringTok{ }\KeywordTok{sample}\NormalTok{(}\KeywordTok{seq\_len}\NormalTok{(}\KeywordTok{NROW}\NormalTok{(cohab\_PopWts)), net\_size, }\OtherTok{TRUE}\NormalTok{, cohab\_PopWts}\OperatorTok{$}\NormalTok{weight)}
      \KeywordTok{set.vertex.attribute}\NormalTok{(nw, }\KeywordTok{names}\NormalTok{(cohab\_PopWts)[}\OperatorTok{{-}}\DecValTok{1}\NormalTok{], cohab\_PopWts[inds,}\OperatorTok{{-}}\DecValTok{1}\NormalTok{])}
\NormalTok{      elapsed \textless{}{-}}\StringTok{ }\KeywordTok{system.time}\NormalTok{(\{}
\NormalTok{        interval \textless{}{-}}\StringTok{ }\DecValTok{500}\OperatorTok{*}\DecValTok{10}\OperatorTok{\^{}}\NormalTok{j; }\ControlFlowTok{if}\NormalTok{ (j}\OperatorTok{==}\DecValTok{3}\NormalTok{) interval \textless{}{-}}\StringTok{ }\OtherTok{NULL}
\NormalTok{        samplesize \textless{}{-}}\StringTok{ }\DecValTok{7500}\NormalTok{; }\ControlFlowTok{if}\NormalTok{ (j}\OperatorTok{==}\DecValTok{3}\NormalTok{) samplesize \textless{}{-}}\StringTok{ }\OtherTok{NULL}
\NormalTok{        effectiveSize \textless{}{-}}\StringTok{ }\DecValTok{64}\NormalTok{; }\ControlFlowTok{if}\NormalTok{ (j}\OperatorTok{\textless{}}\DecValTok{3}\NormalTok{) effectiveSize \textless{}{-}}\StringTok{ }\OtherTok{NULL}
\NormalTok{        fit \textless{}{-}}\StringTok{ }\KeywordTok{ergm}\NormalTok{(nw }\OperatorTok{\textasciitilde{}}\StringTok{ }\NormalTok{edges }\OperatorTok{+}
\StringTok{                    }\KeywordTok{nodefactor}\NormalTok{(}\StringTok{"sex.ident"}\NormalTok{, }\DataTypeTok{levels =} \DecValTok{3}\NormalTok{) }\OperatorTok{+}
\StringTok{                    }\KeywordTok{nodecov}\NormalTok{(}\StringTok{"age"}\NormalTok{) }\OperatorTok{+}
\StringTok{                    }\KeywordTok{nodecov}\NormalTok{(}\StringTok{"agesq"}\NormalTok{) }\OperatorTok{+}
\StringTok{                    }\KeywordTok{nodefactor}\NormalTok{(}\StringTok{"race"}\NormalTok{, }\DataTypeTok{levels =} \DecValTok{{-}5}\NormalTok{) }\OperatorTok{+}
\StringTok{                    }\KeywordTok{nodefactor}\NormalTok{(}\StringTok{"othr.net.deg"}\NormalTok{, }\DataTypeTok{levels =} \DecValTok{{-}1}\NormalTok{) }\OperatorTok{+}
\StringTok{                    }\KeywordTok{nodematch}\NormalTok{(}\StringTok{"race"}\NormalTok{, }\DataTypeTok{diff =} \OtherTok{TRUE}\NormalTok{) }\OperatorTok{+}
\StringTok{                    }\KeywordTok{absdiff}\NormalTok{(}\StringTok{"sqrt.age.adj"}\NormalTok{) }\OperatorTok{+}
\StringTok{                    }\KeywordTok{offset}\NormalTok{(}\KeywordTok{nodematch}\NormalTok{(}\StringTok{"sex"}\NormalTok{, }\DataTypeTok{diff =} \OtherTok{FALSE}\NormalTok{)) }\OperatorTok{+}
\StringTok{                    }\KeywordTok{offset}\NormalTok{(concurrent),}
                    \DataTypeTok{target.stats =}\NormalTok{ cohab\_TargetStats,}
                    \DataTypeTok{offset.coef =} \KeywordTok{c}\NormalTok{(}\OperatorTok{{-}}\OtherTok{Inf}\NormalTok{, }\OperatorTok{{-}}\OtherTok{Inf}\NormalTok{),}
                    \DataTypeTok{eval.loglik =} \OtherTok{FALSE}\NormalTok{,}
                    \DataTypeTok{constraints =} \OperatorTok{\textasciitilde{}}\KeywordTok{bd}\NormalTok{(}\DataTypeTok{maxout =} \DecValTok{1}\NormalTok{) }\OperatorTok{+}\StringTok{ }\KeywordTok{blocks}\NormalTok{(}\DataTypeTok{attr =} \OperatorTok{\textasciitilde{}}\NormalTok{sex, }\DataTypeTok{levels2 =} \KeywordTok{diag}\NormalTok{(}\OtherTok{TRUE}\NormalTok{, }\DecValTok{2}\NormalTok{)),}
                    \DataTypeTok{control =} \KeywordTok{snctrl}\NormalTok{(}\DataTypeTok{MCMC.prop =} \OperatorTok{\textasciitilde{}}\KeywordTok{strat}\NormalTok{(}\DataTypeTok{attr =} \OperatorTok{\textasciitilde{}}\NormalTok{race, }\DataTypeTok{empirical =} \OtherTok{TRUE}\NormalTok{) }\OperatorTok{+}\StringTok{ }\NormalTok{sparse,}
                                     \DataTypeTok{init.method =} \StringTok{"MPLE"}\NormalTok{,}
                                     \DataTypeTok{init.MPLE.samplesize =} \FloatTok{5e7}\NormalTok{,}
                                     \DataTypeTok{MPLE.constraints.ignore =} \OtherTok{TRUE}\NormalTok{,}
                                     \DataTypeTok{MCMLE.effectiveSize =}\NormalTok{ effectiveSize,}
                                     \DataTypeTok{MCMC.burnin =}\NormalTok{ interval,}
                                     \DataTypeTok{MCMC.interval =}\NormalTok{ interval,}
                                     \DataTypeTok{MCMC.samplesize =}\NormalTok{ samplesize,}
                                     \DataTypeTok{parallel =}\NormalTok{ ncores,}
                                     \DataTypeTok{SAN.nsteps =} \FloatTok{5e7}\NormalTok{,}
                                     \DataTypeTok{SAN.prop=}\OperatorTok{\textasciitilde{}}\KeywordTok{strat}\NormalTok{(}\DataTypeTok{attr =} \OperatorTok{\textasciitilde{}}\NormalTok{race, }\DataTypeTok{pmat =}\NormalTok{ cohab\_MixMat) }\OperatorTok{+}\StringTok{ }\NormalTok{sparse))}
\NormalTok{      \})}
\NormalTok{      V4Times[[j }\OperatorTok{+}\StringTok{ }\NormalTok{(i}\DecValTok{{-}1}\NormalTok{)}\OperatorTok{*}\DecValTok{3}\NormalTok{]] \textless{}{-}}\StringTok{ }\NormalTok{elapsed[}\DecValTok{3}\NormalTok{] }\CommentTok{\# save elapsed time}
\NormalTok{    \}}
\NormalTok{  \}}
\NormalTok{\}}
\end{Highlighting}
\end{Shaded}

To test \pkg{ergm} version 3.10, we use the \pkg{callr} package to open
a new \proglang{R} process, install the older version of \pkg{ergm}
along with the related packages \pkg{statnet.common} and \pkg{network},
and run a lengthy example in the new process. To run this code requires
the installation of the \pkg{dplyr} and \pkg{lpSolve} packages, which
were dependencies of \pkg{ergm} 3.10, along with the \pkg{callr}
package.

\begin{Shaded}
\begin{Highlighting}[]
\NormalTok{Version3Code \textless{}{-}}\StringTok{ }\ControlFlowTok{function}\NormalTok{(cohab\_PopWts, cohab\_TargetStats, ncores) \{}
  \CommentTok{\# First, install older versions of three packages including ergm 3.10 to current dir}
\NormalTok{  archv \textless{}{-}}\StringTok{ "https://cran.r{-}project.org/src/contrib/Archive/"}
  \KeywordTok{install.packages}\NormalTok{(}\KeywordTok{paste}\NormalTok{(archv, }\StringTok{"statnet.common/statnet.common\_4.3.0.tar.gz"}\NormalTok{, }\DataTypeTok{sep=}\StringTok{""}\NormalTok{),}
                   \DataTypeTok{repos =} \OtherTok{NULL}\NormalTok{, }\DataTypeTok{type =} \StringTok{"source"}\NormalTok{, }\DataTypeTok{lib=}\StringTok{"."}\NormalTok{)}
  \KeywordTok{library}\NormalTok{(statnet.common, }\DataTypeTok{lib=}\StringTok{"."}\NormalTok{)}
  \KeywordTok{install.packages}\NormalTok{(}\KeywordTok{paste}\NormalTok{(archv, }\StringTok{"network/network\_1.15.tar.gz"}\NormalTok{, }\DataTypeTok{sep=}\StringTok{""}\NormalTok{),}
                   \DataTypeTok{repos =} \OtherTok{NULL}\NormalTok{, }\DataTypeTok{type =} \StringTok{"source"}\NormalTok{, }\DataTypeTok{lib=}\StringTok{"."}\NormalTok{)}
  \KeywordTok{library}\NormalTok{(network, }\DataTypeTok{lib=}\StringTok{"."}\NormalTok{)}
  \KeywordTok{install.packages}\NormalTok{(}\KeywordTok{paste}\NormalTok{(archv, }\StringTok{"ergm/ergm\_3.10.4.tar.gz"}\NormalTok{, }\DataTypeTok{sep=}\StringTok{""}\NormalTok{),}
                   \DataTypeTok{repos =} \OtherTok{NULL}\NormalTok{, }\DataTypeTok{type =} \StringTok{"source"}\NormalTok{, }\DataTypeTok{lib=}\StringTok{"."}\NormalTok{)}
  \KeywordTok{library}\NormalTok{(ergm, }\DataTypeTok{lib=}\StringTok{"."}\NormalTok{)}
\NormalTok{  net\_size \textless{}{-}}\StringTok{ }\FloatTok{5e4}
  \KeywordTok{set.seed}\NormalTok{(}\DecValTok{1}\NormalTok{)}
\NormalTok{  nw \textless{}{-}}\StringTok{ }\KeywordTok{network.initialize}\NormalTok{(net\_size, }\DataTypeTok{directed =} \OtherTok{FALSE}\NormalTok{)}
\NormalTok{  inds \textless{}{-}}\StringTok{ }\KeywordTok{sample}\NormalTok{(}\KeywordTok{seq\_len}\NormalTok{(}\KeywordTok{NROW}\NormalTok{(cohab\_PopWts)), net\_size, }\OtherTok{TRUE}\NormalTok{, cohab\_PopWts}\OperatorTok{$}\NormalTok{weight)}
  \KeywordTok{set.vertex.attribute}\NormalTok{(nw, }\KeywordTok{names}\NormalTok{(cohab\_PopWts)[}\OperatorTok{{-}}\DecValTok{1}\NormalTok{], cohab\_PopWts[inds,}\OperatorTok{{-}}\DecValTok{1}\NormalTok{])}
\NormalTok{  attrib\_mat \textless{}{-}}\StringTok{ }\KeywordTok{matrix}\NormalTok{(}\OtherTok{FALSE}\NormalTok{, }\DataTypeTok{nrow =}\NormalTok{ net\_size, }\DataTypeTok{ncol =} \DecValTok{2}\NormalTok{)}
\NormalTok{  attrib\_mat[nw }\OperatorTok{\%v\%}\StringTok{ "sex"} \OperatorTok{==}\StringTok{ "F"}\NormalTok{, }\DecValTok{1}\NormalTok{] \textless{}{-}}\StringTok{ }\OtherTok{TRUE}
\NormalTok{  attrib\_mat[nw }\OperatorTok{\%v\%}\StringTok{ "sex"} \OperatorTok{==}\StringTok{ "M"}\NormalTok{, }\DecValTok{2}\NormalTok{] \textless{}{-}}\StringTok{ }\OtherTok{TRUE}
\NormalTok{  maxout\_mat \textless{}{-}}\StringTok{ }\KeywordTok{matrix}\NormalTok{(}\DecValTok{0}\NormalTok{, }\DataTypeTok{nrow =}\NormalTok{ net\_size, }\DataTypeTok{ncol =} \DecValTok{2}\NormalTok{)}
\NormalTok{  maxout\_mat[nw }\OperatorTok{\%v\%}\StringTok{ "sex"} \OperatorTok{==}\StringTok{ "F"}\NormalTok{, }\DecValTok{2}\NormalTok{] \textless{}{-}}\StringTok{ }\DecValTok{1}
\NormalTok{  maxout\_mat[nw }\OperatorTok{\%v\%}\StringTok{ "sex"} \OperatorTok{==}\StringTok{ "M"}\NormalTok{, }\DecValTok{1}\NormalTok{] \textless{}{-}}\StringTok{ }\DecValTok{1}
\NormalTok{  V3Times \textless{}{-}}\StringTok{ }\KeywordTok{list}\NormalTok{()}
\NormalTok{  trial \textless{}{-}}\StringTok{ }\DecValTok{1}
  \ControlFlowTok{for}\NormalTok{(interval }\ControlFlowTok{in} \KeywordTok{c}\NormalTok{(}\FloatTok{1e5}\NormalTok{, }\FloatTok{1e6}\NormalTok{)) \{}
  \CommentTok{\# calculate fit without bounded degree contraints}
\NormalTok{    elapsed \textless{}{-}}\StringTok{ }\KeywordTok{system.time}\NormalTok{ (\{}
\NormalTok{      fit \textless{}{-}}\StringTok{ }\KeywordTok{ergm}\NormalTok{(nw }\OperatorTok{\textasciitilde{}}\StringTok{ }\NormalTok{edges }\OperatorTok{+}
\StringTok{                   }\KeywordTok{nodefactor}\NormalTok{(}\StringTok{"sex.ident"}\NormalTok{, }\DataTypeTok{levels =} \DecValTok{3}\NormalTok{) }\OperatorTok{+}
\StringTok{                   }\KeywordTok{nodecov}\NormalTok{(}\StringTok{"age"}\NormalTok{) }\OperatorTok{+}
\StringTok{                   }\KeywordTok{nodecov}\NormalTok{(}\StringTok{"agesq"}\NormalTok{) }\OperatorTok{+}
\StringTok{                   }\KeywordTok{nodefactor}\NormalTok{(}\StringTok{"race"}\NormalTok{, }\DataTypeTok{levels =} \DecValTok{{-}5}\NormalTok{) }\OperatorTok{+}
\StringTok{                   }\KeywordTok{nodefactor}\NormalTok{(}\StringTok{"othr.net.deg"}\NormalTok{, }\DataTypeTok{levels =} \DecValTok{{-}1}\NormalTok{) }\OperatorTok{+}
\StringTok{                   }\KeywordTok{nodematch}\NormalTok{(}\StringTok{"race"}\NormalTok{, }\DataTypeTok{diff =} \OtherTok{TRUE}\NormalTok{) }\OperatorTok{+}
\StringTok{                   }\KeywordTok{absdiff}\NormalTok{(}\StringTok{"sqrt.age.adj"}\NormalTok{) }\OperatorTok{+}
\StringTok{                   }\KeywordTok{offset}\NormalTok{(}\KeywordTok{nodematch}\NormalTok{(}\StringTok{"sex"}\NormalTok{, }\DataTypeTok{diff =} \OtherTok{FALSE}\NormalTok{)) }\OperatorTok{+}
\StringTok{                   }\KeywordTok{offset}\NormalTok{(concurrent),}
                   \DataTypeTok{target.stats =}\NormalTok{ cohab\_TargetStats,}
                   \DataTypeTok{offset.coef =} \KeywordTok{c}\NormalTok{(}\OperatorTok{{-}}\OtherTok{Inf}\NormalTok{, }\OperatorTok{{-}}\OtherTok{Inf}\NormalTok{),}
                   \DataTypeTok{eval.loglik =} \OtherTok{FALSE}\NormalTok{,}
                   \DataTypeTok{control =} \KeywordTok{control.ergm}\NormalTok{(}\DataTypeTok{init.method =} \StringTok{"MPLE"}\NormalTok{,}
                                          \DataTypeTok{MCMC.burnin =}\NormalTok{ interval,}
                                          \DataTypeTok{MCMC.interval =}\NormalTok{ interval,}
                                          \DataTypeTok{MCMC.samplesize =} \DecValTok{7500}\NormalTok{,}
                                          \DataTypeTok{parallel =}\NormalTok{ ncores,}
                                          \DataTypeTok{MCMLE.maxit =} \DecValTok{1000}\NormalTok{,}
                                          \DataTypeTok{MCMLE.termination =} \StringTok{"none"}\NormalTok{,}
                                          \DataTypeTok{SAN.control =} \KeywordTok{control.san}\NormalTok{(}\DataTypeTok{SAN.nsteps =} \FloatTok{1e4}\NormalTok{)))}
\NormalTok{    \})}
\NormalTok{    V3Times[[trial]] \textless{}{-}}\StringTok{ }\NormalTok{elapsed[}\DecValTok{3}\NormalTok{]}
\NormalTok{    trial \textless{}{-}}\StringTok{ }\NormalTok{trial }\OperatorTok{+}\StringTok{ }\DecValTok{1}
    \CommentTok{\# calculate fit with bounded degree contraints}
\NormalTok{    elapsed \textless{}{-}}\StringTok{ }\KeywordTok{system.time}\NormalTok{(\{}
\NormalTok{      fit \textless{}{-}}\StringTok{ }\KeywordTok{ergm}\NormalTok{(nw }\OperatorTok{\textasciitilde{}}\StringTok{ }\NormalTok{edges }\OperatorTok{+}
\StringTok{                   }\KeywordTok{nodefactor}\NormalTok{(}\StringTok{"sex.ident"}\NormalTok{, }\DataTypeTok{levels =} \DecValTok{3}\NormalTok{) }\OperatorTok{+}
\StringTok{                   }\KeywordTok{nodecov}\NormalTok{(}\StringTok{"age"}\NormalTok{) }\OperatorTok{+}
\StringTok{                   }\KeywordTok{nodecov}\NormalTok{(}\StringTok{"agesq"}\NormalTok{) }\OperatorTok{+}
\StringTok{                   }\KeywordTok{nodefactor}\NormalTok{(}\StringTok{"race"}\NormalTok{, }\DataTypeTok{levels =} \DecValTok{{-}5}\NormalTok{) }\OperatorTok{+}
\StringTok{                   }\KeywordTok{nodefactor}\NormalTok{(}\StringTok{"othr.net.deg"}\NormalTok{, }\DataTypeTok{levels =} \DecValTok{{-}1}\NormalTok{) }\OperatorTok{+}
\StringTok{                   }\KeywordTok{nodematch}\NormalTok{(}\StringTok{"race"}\NormalTok{, }\DataTypeTok{diff =} \OtherTok{TRUE}\NormalTok{) }\OperatorTok{+}
\StringTok{                   }\KeywordTok{absdiff}\NormalTok{(}\StringTok{"sqrt.age.adj"}\NormalTok{) }\OperatorTok{+}
\StringTok{                   }\KeywordTok{offset}\NormalTok{(}\KeywordTok{nodematch}\NormalTok{(}\StringTok{"sex"}\NormalTok{, }\DataTypeTok{diff =} \OtherTok{FALSE}\NormalTok{)) }\OperatorTok{+}
\StringTok{                   }\KeywordTok{offset}\NormalTok{(concurrent),}
                   \DataTypeTok{target.stats =}\NormalTok{ cohab\_TargetStats,}
                   \DataTypeTok{offset.coef =} \KeywordTok{c}\NormalTok{(}\OperatorTok{{-}}\OtherTok{Inf}\NormalTok{, }\OperatorTok{{-}}\OtherTok{Inf}\NormalTok{),}
                   \DataTypeTok{eval.loglik =} \OtherTok{FALSE}\NormalTok{,}
                   \DataTypeTok{constraints =} \OperatorTok{\textasciitilde{}}\KeywordTok{bd}\NormalTok{(}\DataTypeTok{attribs =}\NormalTok{ attrib\_mat, }\DataTypeTok{maxout =}\NormalTok{ maxout\_mat),}
                   \DataTypeTok{control =} \KeywordTok{control.ergm}\NormalTok{(}\DataTypeTok{init.method =} \StringTok{"MPLE"}\NormalTok{,}
                                          \DataTypeTok{MCMC.burnin =}\NormalTok{ interval,}
                                          \DataTypeTok{MCMC.interval =}\NormalTok{ interval,}
                                          \DataTypeTok{MCMC.samplesize =} \DecValTok{7500}\NormalTok{,}
                                          \DataTypeTok{parallel =}\NormalTok{ ncores,}
                                          \DataTypeTok{MCMLE.maxit =} \DecValTok{1000}\NormalTok{,}
                                          \DataTypeTok{MCMLE.termination =} \StringTok{"none"}\NormalTok{,}
                                          \DataTypeTok{SAN.control =} \KeywordTok{control.san}\NormalTok{(}\DataTypeTok{SAN.nsteps =} \FloatTok{1e4}\NormalTok{)))}
\NormalTok{    \})}
\NormalTok{    V3Times[[trial]] \textless{}{-}}\StringTok{ }\NormalTok{elapsed[}\DecValTok{3}\NormalTok{] }\CommentTok{\# save elapsed time}
\NormalTok{    trial \textless{}{-}}\StringTok{ }\NormalTok{trial }\OperatorTok{+}\StringTok{ }\DecValTok{1}\NormalTok{;}
\NormalTok{  \}}
\NormalTok{  V3Times}
\NormalTok{\}}
\ControlFlowTok{if}\NormalTok{ (RunMode }\OperatorTok{==}\StringTok{ "Large"}\NormalTok{) \{ }\CommentTok{\# There is no "Small" option here; "Small" works like "Skip"}
  \KeywordTok{library}\NormalTok{(dplyr) }\CommentTok{\# Both dplyr and lpSolve are dependencies of ergm 3.10}
  \KeywordTok{library}\NormalTok{(lpSolve)}
  \KeywordTok{library}\NormalTok{(callr)}
\NormalTok{  V3Times \textless{}{-}}\StringTok{ }\KeywordTok{r}\NormalTok{(Version3Code, }\DataTypeTok{args=}\KeywordTok{list}\NormalTok{(}\DataTypeTok{cohab\_PopWts=}\NormalTok{cohab\_PopWts,}
                                       \DataTypeTok{cohab\_TargetStats=}\NormalTok{cohab\_TargetStats, }\DataTypeTok{ncores=}\NormalTok{ncores))}
\NormalTok{\}}
\end{Highlighting}
\end{Shaded}

\begin{table}

\caption{\label{tab:Estimationtable}Model fit times for various {\bf ergm} versions and settings using fixed short, fixed long, and adaptive MCMC intervals.}
\centering
\begin{tabular}[t]{lrrr}
\toprule
  & Short & Long & Adaptive\\
\midrule
3.10 without \texttt{bd} & 22.64 hours (1e5 interval) & 4.41 hours (1e6 interval) & N/A\\
3.10 with \texttt{bd} & 5.72 hours (1e5 interval) & 1.3 hours (1e6 interval) & N/A\\
4.1 & 281.5 seconds (5e3 interval) & 658.2 seconds (5e4 interval) & 569 seconds\\
\bottomrule
\end{tabular}
\end{table}

\hypertarget{discussion}{%
\section{Discussion}\label{discussion}}

This paper describes and tests the many changes in the \pkg{ergm}
package since version 2.1 was released concurrently with Hunter et al.
(2008) that influence the computing efficiency of the various Monte
Carlo-based algorithms upon which it depends. Among other things, we
demonstrate that the computing algorithms have improved by up to several
orders of magnitude; coupled with the concomitant increase of processor
speed, these developments enable \pkg{ergm} users to model networks of a
size that was infeasible a decade ago. Furthermore, \pkg{ergm} the many
related packages in the \statnet{} suite for \proglang{R} (\proglang{R}
Core Team, 2021) are undergoing continual development, and we intend
that this trend will continue.

\hypertarget{acknowledgments}{%
\section*{Acknowledgments}\label{acknowledgments}}
\addcontentsline{toc}{section}{Acknowledgments}

Many individuals have contributed code for version 4 of \pkg{ergm},
particularly Mark Handcock, who wrote most of the code upon which
missing data inference and diagnostics are based, and Michał Bojanowski,
who produced the \texttt{predict} method, among many other contributions
by both of them. Carter Butts is the main developer of the \pkg{network}
package, upon which \pkg{ergm} depends; in addition, he provided
numerous suggestions for computational improvements and new terms, and
provided numerous helpful comments about this manuscript. Skye
Bender-deMoll wrote a vignette that automatically cross-references
\texttt{ergm} model terms, Joyce Cheng wrote the dynamic documentation
system and miscellaneous enhancements, and Christian Schmid contributed
code improving MPLE standard error estimation. Other important
contributors are Steven Goodreau, Ayn Leslie-Cook, Li Wang, and Kirk Li.
We are grateful to all these individuals as well as the many users of
\pkg{ergm} who have aided the package's development through the many
questions and suggestions they have posed over the years.

\hypertarget{references}{%
\section*{References}\label{references}}
\addcontentsline{toc}{section}{References}

\hypertarget{refs}{}
\begin{cslreferences}
\leavevmode\hypertarget{ref-Asuncion2010}{}%
Asuncion, A., Liu, Q., Ihler, A., \& Smyth, P. (2010). Learning with
blocks: Composite likelihood and contrastive divergence. In Y. W. Teh \&
M. Titterington (Eds.), \emph{Proceedings of the thirteenth
international conference on artificial intelligence and statistics}
(Vol. 9, pp. 33--40). PMLR.
\url{http://proceedings.mlr.press/v9/asuncion10a.html}

\leavevmode\hypertarget{ref-R2021}{}%
\proglang{R} Core Team. (2021). \emph{R: A language and environment for
statistical computing}. R Foundation for Statistical Computing.
\url{http://www.R-project.org/}

\leavevmode\hypertarget{ref-Gelman1998}{}%
Gelman, A., \& Meng, X.-L. (1998). Simulating normalizing constants:
From importance sampling to bridge sampling to path sampling.
\emph{Statistical Science}, \emph{13}, 163--185.

\leavevmode\hypertarget{ref-Ge91b}{}%
Geweke, J. (1991). \emph{Bayesian statistics 4} (J. M. Bernado, J. O.
Berger, A. P. Dawid, \& A. F. M. Smith, Eds.). Federal Reserve Bank of
Minneapolis, Research Department Minneapolis, MN, USA.

\leavevmode\hypertarget{ref-handcock2003}{}%
Handcock, M. S. (2003). \emph{Assessing degeneracy in statistical models
of social networks}. University of Washington.
\url{https://csss.uw.edu/files/working-papers/2003/wp39.pdf}

\leavevmode\hypertarget{ref-HuHu12i}{}%
Hummel, R. M., Hunter, D. R., \& Handcock, M. S. (2012). Improving
simulation-based algorithms for fitting ERGMs. \emph{Journal of
Computational and Graphical Statistics}, \emph{21}(4), 920--939.
\url{https://doi.org/10.1080/10618600.2012.679224}

\leavevmode\hypertarget{ref-HuHa06i}{}%
Hunter, D. R., \& Handcock, M. S. (2006). Inference in curved
exponential family models for networks. \emph{Journal of Computational
and Graphical Statistics}, \emph{15}(3), 565--583.
\url{https://doi.org/10.1198/106186006x133069}

\leavevmode\hypertarget{ref-HuHa08e}{}%
Hunter, D. R., Handcock, M. S., Butts, C. T., Goodreau, S. M., \&
Morris, M. (2008). ergm: A package to fit, simulate and diagnose
exponential-family models for networks. \emph{Journal of Statistical
Software}, \emph{24}(3), 1--29.
\url{https://doi.org/10.18637/jss.v024.i03}

\leavevmode\hypertarget{ref-Kr12e}{}%
Krivitsky, P. N. (2012). Exponential-family random graph models for
valued networks. \emph{Electronic Journal of Statistics}, \emph{6},
1100--1128. \url{https://doi.org/10.1214/12-EJS696}

\leavevmode\hypertarget{ref-Kr17u}{}%
Krivitsky, P. N. (2017). Using contrastive divergence to seed Monte
Carlo MLE for exponential-family random graph models.
\emph{Computational Statistics \& Data Analysis}, \emph{107}, 149--161.
\url{https://doi.org/10.1016/j.csda.2016.10.015}

\leavevmode\hypertarget{ref-KrBu17e}{}%
Krivitsky, P. N., \& Butts, C. T. (2017). Exponential-family random
graph models for rank-order relational data. \emph{Sociological
Methodology}, \emph{47}(1), 68--112.
\url{https://doi.org/10.1177/0081175017692623}

\leavevmode\hypertarget{ref-ergm4a}{}%
Krivitsky, P. N., Hunter, D. R., Morris, M., \& Klumb, C. (2022).
\emph{ergm 4: New features}. \url{https://arxiv.org/abs/2106.04997v2}

\leavevmode\hypertarget{ref-krivitsky2022}{}%
Krivitsky, P. N., Kuvelkar, A. R., \& Hunter, D. R. (2022).
\emph{Likelihood-based inference for exponential-family random graph
models via linear programming}. \url{https://arxiv.org/abs/2202.03572v1}

\leavevmode\hypertarget{ref-Meng1996}{}%
Meng, X.-L., \& Wong, W. H. (1996). Simulating ratios of normalizing
constants via a simple identity: A theoretical exploration.
\emph{Statistica Sinica}, \emph{6}, 831--860.

\leavevmode\hypertarget{ref-MoHa08spec}{}%
Morris, M., Handcock, M. S., \& Hunter, D. R. (2008). Specification of
exponential-family random graph models: Terms and computational aspects.
\emph{Journal of Statistical Software}, \emph{24}(4), 1--24.
\url{https://doi.org/10.18637/jss.v024.i04}

\leavevmode\hypertarget{ref-nsfg}{}%
National Center for Health Statistics. (2020). \emph{2006--2015 national
survey of family growth}. \url{https://www.cdc.gov/nchs/nsfg/index.htm}

\leavevmode\hypertarget{ref-PlBe06c}{}%
Plummer, M., Best, N., Cowles, K., \& Vines, K. (2006). CODA:
Convergence diagnosis and output analysis for MCMC. \emph{R News},
\emph{6}(1), 7--11. \url{https://journal.r-project.org/archive/}

\leavevmode\hypertarget{ref-SchmidDesmarais17}{}%
Schmid, C. S., \& Desmarais, B. A. (2017). Exponential random graph
models with big networks: Maximum pseudolikelihood estimation and the
parametric bootstrap. \emph{2017 IEEE International Conference on Big
Data (Big Data)}, 116--121.
\url{https://doi.org/10.1109/bigdata.2017.8257919}

\leavevmode\hypertarget{ref-schmid2020}{}%
Schmid, C. S., \& Hunter, D. R. (2020). \emph{Improving ERGM starting
values using simulated annealing}.
\url{https://arxiv.org/abs/2009.01202}

\leavevmode\hypertarget{ref-schmid2021}{}%
Schmid, C. S., \& Hunter, D. R. (2021). \emph{Accounting for model
misspecification when using pseudolikelihood for ERGMs}.

\leavevmode\hypertarget{ref-schweinberger2020}{}%
Schweinberger, M., Krivitsky, P. N., Butts, C. T., \& Stewart, J. R.
(2020). Exponential-family models of random graphs: Inference in finite,
super and infinite population scenarios. \emph{Statistical Science},
\emph{35}(4), 627--662. \url{https://doi.org/10.1214/19-STS743}

\leavevmode\hypertarget{ref-Sn02m}{}%
Snijders, T. A. B. (2002). Markov chain Monte Carlo estimation of
exponential random graph models. \emph{Journal of Social Structure},
\emph{3}(2).
\url{https://www.cmu.edu/joss/content/articles/volume3/Snijders.pdf}

\leavevmode\hypertarget{ref-VaFl19m}{}%
Vats, D., Flegal, J. M., \& Jones, G. L. (2019). Multivariate output
analysis for Markov chain Monte Carlo. \emph{Biometrika}, \emph{106}(2),
321--337. \url{https://doi.org/10.1093/biomet/asz002}
\end{cslreferences}

\end{document}